\documentclass[prx,aps,twocolumn,showpacs,superscriptaddress,nofootinbib]{revtex4-1}
\usepackage[colorlinks=true,citecolor=blue,linkcolor=magenta]{hyperref}
\usepackage{graphicx}
\usepackage{dcolumn}
\usepackage{bm}
\usepackage[bbgreekl]{mathbbol}
\usepackage{amsmath,dsfont}
\usepackage[normalem]{ulem}
\usepackage{color}

\newcommand{\bra}[1]{\langle #1|}
\newcommand{\ket}[1]{|#1\rangle}
\newcommand{\appropto}{\mathrel{\vcenter{
  \offinterlineskip\halign{\hfil$##$\cr
    \propto\cr\noalign{\kern2pt}\sim\cr\noalign{\kern-2pt}}}}}
\newcommand{\doublearrow}[1]{\overset{\text{\tiny$\leftrightarrow$}}{#1}}

\newcommand{\ex}[1]{\langle #1 \rangle}
\newcommand{\tbl}[1]{#1}
\newcommand{\tblue}[1]{\textcolor{black}{#1}}

\begin{document}
\onecolumngrid
\title{Critical open-system dynamics in a one-dimensional optical lattice clock}

\author{Lo\"{i}c Henriet}
\address{ICFO-Institut de Ciencies Fotoniques, The Barcelona Institute of Science and Technology, 08860 Castelldefels (Barcelona), Spain}

\author{James S. Douglas}
\address{ICFO-Institut de Ciencies Fotoniques, The Barcelona Institute of Science and Technology, 08860 Castelldefels (Barcelona), Spain}

\author{Darrick E. Chang}
\address{ICFO-Institut de Ciencies Fotoniques, The Barcelona Institute of Science and Technology, 08860 Castelldefels (Barcelona), Spain}
\address{ICREA-Instituci\'{o} Catalana de Recerca i Estudis Avan\c{c}ats, 08015 Barcelona, Spain}

\author{Andreas Albrecht}
\address{ICFO-Institut de Ciencies Fotoniques, The Barcelona Institute of Science and Technology, 08860 Castelldefels (Barcelona), Spain}

\date{\today}
\begin{abstract}
There have been concerted efforts in recent years to realize the next generation of clocks using alkaline earth atoms in an optical lattice. Assuming that the atoms are independent, such a clock would benefit from a $\sqrt{N}$ enhancement in its stability, associated with the improved signal-to-noise ratio of a large atom number $N$. An interesting question, however, is what type of atomic interactions might affect the clock dynamics, and whether these interactions are deleterious or could even be beneficial. In this work, we investigate the effect of dipole-dipole interactions, in which atoms excited during the clock protocol emit and re-absorb photons. Taking a simple system consisting of a 1D atomic array, we find that dipole-dipole interactions in fact result in \tbl{an open quantum system exhibiting critical dynamics}, as a set of collective excitations acquires a decay rate approaching zero in the thermodynamic limit due to subradiance. A first consequence is that the decay of atomic excited population at long times exhibits a slow power-law behavior, instead of the exponential expected for non-interacting atoms. We also find that excitations among the atoms exhibit fermionic spatial correlations at long times, due to the microscopic properties of the multi-excitation subradiant states. Interestingly, these properties cannot be captured by mean-field dynamics, suggesting {the} strongly interacting nature of this system. We finally characterize the time-dependent frequency shift in the atomic frequency measurement, and find that it is dominated by the interaction energy of subradiant states at long times. Furthermore, we show that the \tbl{decay of the} clock signal displays at long times a non-exponential behavior, which \tbl{might be useful} to improve the uncertainty limit with which the atomic frequency can be resolved. \tblue{We attribute the lack of robust power-law dynamics for the clock signal to an effective many-body dephasing caused by purely coherent interactions.} 
\end{abstract}
\maketitle

\section{Introduction}

Optical lattice clocks involving a large number of atoms confined in an optical lattice have recently reached unprecedented {performance levels} \,\cite{Bloom2014,Nicholson2015,Campbell17,takamoto05} due to the {combination of the narrow atomic transitions used and the signal-to-noise enhancement associated with atom number}. Moreover, significant progress has been made toward managing systematic errors that would manifest themselves even at the single-atom level \,\cite{ludlow15,Katori03,Ludlow08,Lemke09}. These experiments now operate near a regime where {photon-mediated dipole-dipole} interactions between atoms can become a potential limit to their accuracy\,\cite{Campbell17}. Understanding the effect of such interactions is a {challenging problem involving the dynamics of an open, many-body system, which has previously only been addressed} exactly for small atom number \,\cite{Ostermann13,Maier14}, or perturbatively at short time scales \,\cite{darrick04} or in mean field approaches \,\cite{kramer16}.

In regular atomic arrays, dipole-dipole interactions significantly affect the atomic properties. \tblue{For example, the spatial ordering naturally gives rise to very strong constructive or destructive interference between the emission coming from different atoms. This manifests itself in the emergence of pronounced sub- and superradiance, a prolongation or shortening of the lifetime of a collective atomic excitation as compared to independent atoms.} These properties have recently been explored in one (1D) \,\cite{Plankensteiner15,Bettles16a,Kornovan16,Sutherland16,asenjo17} and two dimensions (2D) \,\cite{Jenkins12,Bettles16b,Facchinetti16,asenjo17,shahmoon17}, where subradiant modes with a single excitation coherently shared among the atoms acquire an intuitive interpretation in terms of optically guided modes. 
In that case, the decay rate {is attributable to scattering into radiation modes at the system boundaries, and decreases polynomially with system size}. The study of subradiant modes with multiple excitations reveals an interesting many-body structure \,\cite{asenjo17}, where excitations obey an effective Pauli exclusion principle and ``fermionize''.

Motivated by the new intuition provided by arrays, here we analyze in a non-perturbative fashion the behavior of a lattice clock composed of a 1D chain of atoms. We choose such a geometry because it enables a non-perturbative analysis and extrapolation to the thermodynamic limit, which could not be done in previous studies of dipole-dipole interactions in clocks \cite{darrick04,Ostermann13,Maier14,kramer16}. Our analysis provides a number of intriguing results. \tblue{Most importantly, the decay rate of subradiant states approaching zero with increasing atom number directly implies that the characteristic time scale for (exponential) relaxation of the system to its equilibrium state can become infinite (so-called "closing of the Liouvillian gap").} This property \tbl{is known to facilitate a critical slowing down of dynamics in a number of open systems}\,\cite{Cai13,medvedyeva14,rota17}. More specifically, \tbl{in our system,} an initially highly excited state reveals a power-law decay of the excited-state \tbl{population at long times}, with a seemingly robust scaling exponent in the thermodynamic limit. The decay process also leads to a smooth, quasi-uniform population of subradiant states at long times, a crucial prerequisite for the observation of an algebraic decay. \tbl{Furthermore, the accumulation of population in subradiant manifolds is accompanied by a build-up of fermionic spatial correlations between excitations. While these effects are quite general, we then proceed to examine a Ramsey-type excitation sequence, as is relevant to optical clocks. Here, we find that subradiant states generate a time-dependent shift in the apparent atomic resonance frequency. } The decay of the clock signal generally exhibits a slow-down, which might be beneficial in improving the clock sensitivity beyond standard quantum limits. However, this slow-down is seen numerically to lack universal behavior. \tblue{In particular, unlike the excited population, the clock signal depends on individual atomic coherences, and we attribute the deviation from a power law decay to an effective many-body dephasing arising from coherent dipole-dipole interactions.}

The paper is structured as follows. In Sec.\,\ref{Sect2}, we introduce the theoretical framework to treat the effect of dipole-dipole interactions in a 1D atomic chain in free space, along with its collective decay properties. Besides arrangements in free space, we also consider a ``toy model'' of atoms coupled through a 1D waveguide. The latter platform reveals similar collective emission properties. Moreover, it can be simulated in the framework of Matrix Product States (MPS), which enables the study of the many-body dynamics for larger system sizes. Numerical results of the decay dynamics are presented in Sec.\,\ref{sec:decay_dynamics}, and reveal a characteristic  power-law region at long times. In Sec.\,\ref{sec_jumppic}, we {introduce a semi-classical rate equation, which enables one to understand the smooth build-up of population in subradiant states that is needed for power-law behavior to emerge}. This approach is further justified by the study of the Liouvillian eigenstructure in Sec.\,\ref{sec:liouvillian}. We then turn in Sec. \ref{sec:clock} to the consequences of these findings for the time evolution of a 1D lattice clock. Finally, \tbl{while a simulation of the full decay dynamics of a large 3D lattice is unfeasible, in Sec.\,\ref{sec:3D} we demonstrate that such a system also exhibits a closing of the Liouvillian gap already in the single-excitation manifold.}

Our results are significant for a number of reasons. First, while the majority of our analysis is restricted to 1D, the essential statement -- that lattice clocks \tbl{exhibit a closing of the Liouvillian gap} -- is true even in 3D, and strongly suggests that a different set of theoretical tools is needed to properly understand \tbl{the dynamics of such} a system. In particular, it suggests that previous perturbative approaches are not sufficient, nor are exact results on small systems in higher dimensions, a case that is completely dominated by boundary effects. Furthermore, within the context of arrays \,\cite{Plankensteiner15,Bettles16a,Kornovan16,Sutherland16,asenjo17,Jenkins12,Bettles16b,Facchinetti16,asenjo17,shahmoon17}, our work goes beyond the typically studied limit of single excitations, which falls within the regime of linear optics, and sheds light on the entire dynamics of collective excitations and subradiance in the many-body limit. It has previously been shown that exploiting subradiance in arrays enables a significant improvement in performance of quantum \tbl{information} protocols\,\cite{asenjo17, manzoni18} and novel phenomena such as topological edge states\,\cite{perczel17, bettles17} within the single-excitation limit. Our work provides important insights and tools to examine applications and phenomena involving many excitations. Recent experimental progress now permits the controlled creation of 1D \cite{Endres16} and 2D \cite{Lester15,barredo16} atomic arrays based on optical tweezers. With further improvements toward smaller inter-atomic distances, the direct and systematic observation of the effects described here comes within reach.

\section{Modeling a one-dimensional optical lattice clock}\label{Sect2}

In this section, we first introduce the  standard clock protocol for an \tbl{ideal, non-interacting} system of two-level atoms. We then present a spin-model formalism {that describes how excited atoms in the lattice interact with common electromagnetic field modes, which gives rise both to coherent exchange-type interactions between atoms and collective emission}.  Collective states in regular arrays show decay properties that drastically differ from the case of independent emitters\,\cite{asenjo17}, and we briefly review the main properties at low excitation densities here.

\subsection{\label{sec:lattice_clock}Clock protocol for an ideal, non-interacting system}

\begin{figure}[h!]
\center
\includegraphics[scale=0.35]{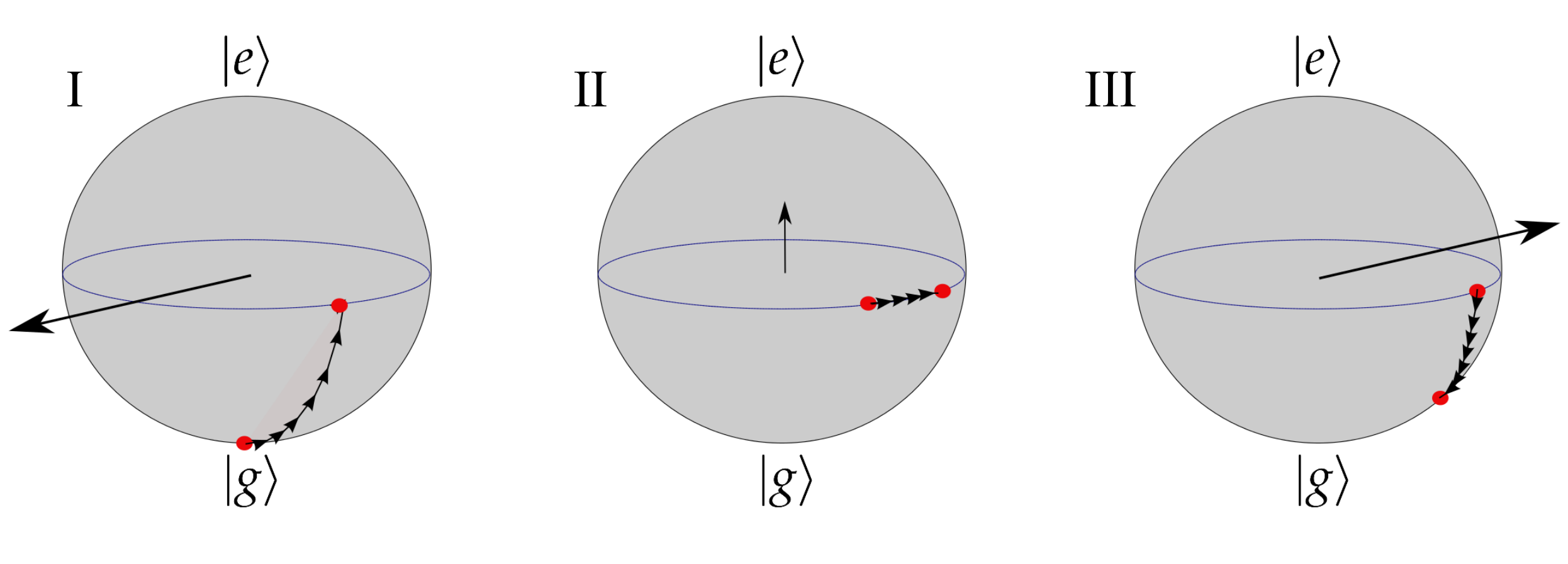}
\caption{\label{FigRamsey} Illustration of the Ramsey sequence on the Bloch sphere, for a single atom. In the first step (I), an initial $\pi/2$ pulse brings the atom in a coherent superposition of ground and excited states. The second step (II) consists of a free evolution under $\mathcal{H}_{\rm indep}$, in which the state vector precesses in the equatorial plane at a rate given by the detuning $\delta=\omega_L-\omega_0$ between the laser frequency and atomic resonance frequency. The last step (III) corresponds to the application of another $\pi/2$ pulse, which maps the precession angle to a measurable population difference between ground and excited states. The black arrow illustrates the instantaneous rotation axis at each step.
}
\end{figure}

{First}, we briefly review the standard Ramsey sequence\,\cite{Ramsey50,Ramsay85} as used in clock protocols for an ideal, non-interacting system of $N$ identical two-level atoms located at positions ${\bf{r}}_n$, with ground and excited states $|g_n \rangle$, $|e_n \rangle$ and transition frequency $\omega_0$. This Ramsey sequence is illustrated in Fig.\,\ref{FigRamsey}. In a first step (I), a global $\pi/2$ pulse -- realized by a strong pulse of an interrogating laser -- initializes the system in a coherent superposition of ground and excited states,
\begin{align}
|\psi_{i}\rangle=\bigotimes_{n=1}^N \frac{|g_n \rangle+e^{i k_L z_n}|e_n \rangle}{\sqrt{2}}.
\label{eq:initial_clock_state}
\end{align}
For concreteness, we assumed ${\bf{k}}_L= k_L  \hat{{\bf{z}}}$ parallel to the z-axis for the wavevector of the laser. During a free evolution of this state for a time $t$, the {ground and excited states of this superposition acquire} a relative phase, which directly serves as the clock signal.  More specifically, for independent atoms, the time evolution of the system density matrix $\rho$ is given by $\dot{\rho}= -(i/\hbar)[\mathcal{H}_{\rm indep} \rho-\rho \mathcal{H}_{\rm indep}^{\dagger}]+\sum_{n}\Gamma_{0}\sigma^n_{ge}\rho \sigma^n_{eg}$, where $\Gamma_0$ is  the spontaneous emission rate of the excited state $|e\rangle$. In the frame rotating with the laser frequency, the Hamiltonian reads $\mathcal{H}_{\rm indep}=\hbar (-\delta -i\Gamma_0/2)\sum_{n} \sigma^n_{ee}$, with $\delta=\omega_L-\omega_0$ the detuning between the laser $\omega_L$ and the atomic transition frequency $\omega_0$. A $\delta$-dependent  relative phase  accumulates during the free-evolution period $t$, which can be viewed as a precession along the equator of the Bloch sphere (II).  After this period, a global $-\pi/2$-pulse (III) maps coherences onto populations for measurement convenience, that is, the angle of precession along the equator of the Bloch sphere is mapped onto a $\delta=\omega_L-\omega_0$ dependent population inversion. 
The $\delta$-dependent signal $S$ associated with that measurement is given by the observable\,\cite{darrick04} $S=-2\Re\left\langle \sum_{n=1}^N e^{ik_L z_n} \sigma_{eg}^n \right\rangle$, where the expectation value is taken right before step (III). For independent atoms this results in $S=-N \cos (\delta t) e^{-\Gamma_0 t/2}$. The determination of the {minimum} of $S$ allows for referencing the laser frequency to the atomic frequency $\delta=0$. Absent any other imperfections, the spontaneous emission enforces an ideal interrogation time of $t\sim \Gamma_0^{-1}$ to avoid substantial signal decay. Repeating such measurements over a total averaging time $T_{\rm avg}$ then allows for the determination of the resonance frequency to {the standard limit of} uncertainty $\Delta\omega \sim \sqrt{\Gamma_0/(NT_{\rm avg})}$.



\subsection{Effective spin-model description}\label{sec_spinmodel}

The Ramsey spectroscopy sensitivity is fundamentally limited by the excited-state spontaneous emission rate $\Gamma_0$, and a $\sqrt{N}$ improvement in signal to noise ratio can be attributed to the \tblue{large number of (presumed independent) atoms comprising the clock}. In a dense atomic system, however, the emission of atoms  becomes collective. {Collective} decay rates then can be much faster (superradiant) or slower (subradiant), as the {fields emitted by the atoms} can interfere {either} constructively {or} destructively. Furthermore, a photon emitted by one excited atom can be coherently re-absorbed by another atom in its ground state, leading to {a dressing of bare atomic energies}. Formally, these processes { can be modeled by integrating out the photonic degrees of freedom from the full atom-light system. This results in an interacting, open spin model,} describing the dynamics of the atomic density matrix $\rho$ as\,\cite{gross82,agarwal70,lehmberg70, akkermans08,asenjo17}
\begin{align}
 \dot{\rho}=\mathcal{L}[\rho]= -(i/\hbar)[\mathcal{H}_{\rm eff}\rho-\rho \mathcal{H}_{\rm eff}^{\dagger}]+\sum_{m,n}\Gamma_{m,n}\,\sigma^m_{ge}\rho \sigma^n_{eg}.
 \label{eq:Liouville_master_equation}
\end{align} 
The Liouvillian super-operator $\mathcal{L}$ is a sum of two terms $\mathcal{L}= \mathcal{K}+\mathcal{J}$, where $\mathcal{K}[\rho]=(\mathcal{H}_{\rm eff} \rho-\rho \mathcal{H}_{\rm eff}^{\dagger})/(i\hbar)$ denotes the coherent-like evolution of the density matrix under the effective Hamiltonian $\mathcal{H}_{\rm eff}$ and $\mathcal{J}[\rho]=\sum_{m,n}\Gamma_{m,n}\sigma^m_{ge}\rho \sigma^n_{eg}$ is the jump, or population-recycling, term. The effective (non-Hermitian) Hamiltonian reads in the rotating frame
\begin{align}
\mathcal{H}_{\rm eff}=-\mu_0 \omega_0^2 \sum_{m,n=1}^N {\bf{p}}_n^{\dagger}\doublearrow{{\bf{G}}}({\bf{r}}_n,{\bf{r}}_m,\omega_0){\bf{p}}_m  \sigma^n_{eg} \sigma^m_{ge}\label{eq:Effective_Hamiltonian}.
\end{align}  
Here, $\mu_0$ is the {vacuum} permeability, ${\bf{p}}_m$ is the dipole {matrix element} of atom $m$, and $\sigma^m_{\alpha \beta}=|\alpha_m \rangle \langle \beta_m |$ defines an operator acting on the internal states $\{\alpha,\beta\}\in \{g,e\}$ of atom $m$. The tensor $\doublearrow{{\bf{G}}}({\bf{r}}_n,{\bf{r}}_m,\omega_0)$ denotes the classical Green's function of the electromagnetic field\,\cite{novotny06,asenjo17}, and the matrix elements $\Gamma_{m,n}=(2\mu_0 \omega_0^2/\hbar){\bf{p}}_n^{\dagger}{\rm Im}\left[\doublearrow{{\bf{G}}}({\bf{r}}_n,{\bf{r}}_m,\omega_0)\right]{\bf{p}}_m $ encode correlated dissipation via coupling to the free-space radiation modes. Throughout this paper, we will focus on one-dimensional atomic chains {oriented along $z$,} with atomic polarization parallel to the chain axis, giving\,\cite{novotny06,asenjo17} ${\bf{p}}^\dagger \doublearrow{{\bf{G}}}({\bf{r}}_n,{\bf{r}}_m,\omega_0){\bf{p}}=(1-ik_0 r) |{\bf{p}}|^2 e^{i k_0 r}/(2\pi k_0^2 r^3)$, where $k_0=\omega_0/c$ and $r=|{\bf{r}}_n-{\bf{r}}_m|$. For a single isolated atom, one recovers the vacuum emission rate $\Gamma_{jj}=\Gamma_{0}$, with $\Gamma_{0}=\omega_0^3 |{\bf{p}}|^2/(3\pi \hbar \epsilon_0 c^3)$. 


\begin{figure}[h!]
\center
\includegraphics[scale=0.6]{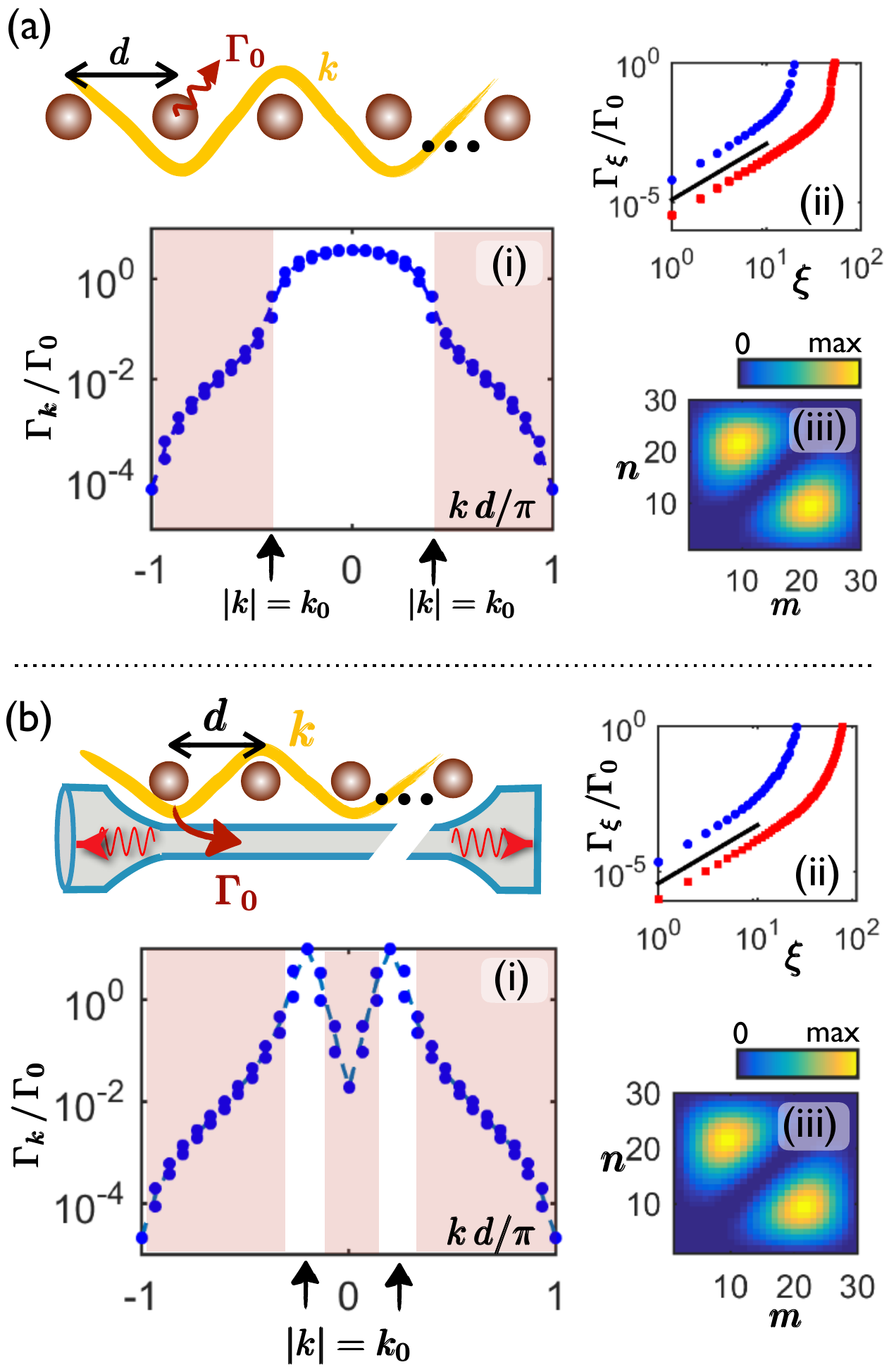}
\caption{\label{Fig1} Atoms forming a periodic chain in (a) free space and (b) along a waveguide. For (i)-(iii) the inter-atomic distance has been chosen such that (a) $d=0.2\lambda_0$ and (b) $k_{\rm 1D}\,d=0.2\pi$ with atom number $N=30$ unless stated otherwise. (i) Single excitation eigenstate decay {rates $\Gamma_k$ (normalized by the single-atom decay rate $\Gamma_0$), as a function of dominant wavevector $k$. The shaded regions indicate values of $k$ where subradiant modes appear due to decoupling of atomic spin waves from electromagnetic radiation. These occur when $|k|>k_0$ in free space and $|k|\neq k_0$ in a waveguide, where $k_0$ is the resonant wavevector}. (ii) Scaling of the subradiant single-excitation decay rates with the numbering index $\xi$ for $N=30$ (blue) and $N=80$ (red). The black line corresponds to a scaling $\Gamma_\xi \sim \xi^2$. (iii) Population {in state} $\ket{e_m,e_n}$, {for} the most subradiant two-excitation eigenstate. }
\end{figure}


\subsection{Collective decay properties}
\label{subsec:subradiant_properties}
In view of the correlated behavior of atoms in dense arrangements, we anticipate the dynamics of a clock to significantly deviate at long times $t\gtrsim \Gamma_0^{-1}$ from independent atom models or mean field theory. The dynamics of the system density matrix is fully characterized by the properties of the Liouvillian superoperator $\mathcal{L}$. It is however very instructive to study first the structure of the effective Hamiltonian $\mathcal{H}_{\rm eff}$, whose properties determine the Liouvillian dynamics to a large extent. In particular, useful insight into the collective dynamics can be obtained from the decay rate spectrum associated with eigenstates of $\mathcal{H}_{\rm eff}$, which we here discuss with a particular focus on long-lived subradiant states {with} decay rates $\Gamma_{\xi}\ll \Gamma_0$. The fact that $\mathcal{H}_{\rm eff}$ commutes with the total number of {excitations} $\hat{n}_e=\sum_n \sigma^n_{ee}$ permits one to identify and characterize eigenstates within each excitation manifold. The properties of eigenstates with low-excitation number ${n}_e\ll N$ have previously been studied in detail in Ref.\,\cite{asenjo17} and we summarize the main results here.

In the single-excitation sector,  diagonalizing the effective Hamiltonian $\mathcal{H}_{\rm eff}$ results in $N$ eigenvectors $|\psi^{(1)}_{\xi}\rangle$, with $\mathcal{H}_{\rm eff}|\psi^{(1)}_{\xi}\rangle=\hbar(\omega^{(1)}_{\xi}-i\Gamma^{(1)}_{\xi}/2)\ket{\psi_\xi^{(1)}}$, where $\omega^{(1)}_{\xi}$ and $\Gamma^{(1)}_{\xi}>0$ represent the energy shifts (relative to the bare frequency $\omega_0$) and decay rates associated with $|\psi^{(1)}_{\xi}\rangle$, respectively. Here, the upper index (1) labels the (single) excitation sector, whereas $1\leq\xi\leq N$ serves as an index for states within the sector. For an infinite chain, the eigenstates of $\mathcal{H}_{\rm eff}$ correspond to spin waves $|\psi^{(1)}_k\rangle=S^{\dagger}_{k} |g\rangle^{\otimes N}$, with $k$ a quantized wavevector within the first Brillouin zone ($|k|\leq \pi/d$) and $S^{\dagger}_{k}=1/\sqrt{N}\sum_n e^{ikdn}\sigma_{\rm eg}^n$. Such states are the exact eigenstates only in the infinite-chain {limit. However}, the spin-wave character persists at finite large $N$ and one can still assign a dominant wavevector $k$ to each eigenstate {by looking for the peak value in the Fourier transform of the real-space atomic excitation amplitudes. In other words, single-excitation eigenstates $\ket{\psi_{k}^{(1)}}$ can be indexed unambiguously by $k$ in place of $\xi$.} We plot in Fig.\,\ref{Fig1}\,(a)\,(i) the collective decay rates $\Gamma^{(1)}_k$ of the eigenstates $|\psi^{(1)}_{k}\rangle=\sum_{n=1}^N c_n^{k}\ket{e_n}$, for a fixed number of atoms $N=30$. One finds that for $|k|>k_0=\omega_0/c$ the decay rate associated with the eigenstate $|\psi^{(1)}_{k}\rangle$ is greatly reduced as compared to the independent emission rate $\Gamma_0$. 

This behavior can be understood by \tbl{considering} the electromagnetic field ${\bf{E}}({\bf{r}})$ generated by such spin waves of wavevector $k$. The field can be expanded in a plane wave basis, with discrete translational invariance ensuring that ${\bf{E}}({\bf{r}})=\sum_{g,{\bf{k}}_{\perp}} {\bf{E}}_{g,{\bf{k}}_{\perp}} e^{i(k+g)z+{\bf{k}}_{\perp}.{\bf{r}}_{\perp}}$ only has axial wavevectors equal to $k$, up to a reciprocal lattice vector $g$. At the same time, the wave equation requires that $(k+g)^2+{\bf{k}}_{\perp}^2=(\omega/c)^2$. Thus, when $|k|>k_0$, the transverse wavevector is purely imaginary. The resulting field cannot radiate away energy and instead is evanescently confined to the chain in the transverse direction, in complete analogy with a fiber guided mode. For {such states to exist}, $k_0$ must be within the first Brillouin zone, which leads to the requirement that the inter-atomic distance $d$ must be smaller than half the wavelength of the atomic transition, $d<\lambda_0/2$. For a finite chain, this mechanism leaves scattering through the ends of the chain as the only decay channel, and explains the greatly reduced decay rates of these excitations. In this case, the most subradiant states show a smooth distribution of decay rates whose minimum value approaches zero in the thermodynamic limit $N\to \infty$. Ordering the eigenstates by increasing decay rates, i.e. from $\xi=1$ for the most subradiant to $\xi=N$ for the most radiant, one finds more specifically that the most subradiant modes are characterized by a decay rate $\Gamma^{(1)}_{\xi} \ll \Gamma_{0}$ {that} scales with atom number as $\Gamma^{(1)}_{\xi}/\Gamma_{0}\propto \xi^2/N^3$. This is illustrated in Fig.\,\ref{Fig1} (a) (ii), where we plot $\Gamma^{(1)}_{\xi} / \Gamma_{0}$ with respect to $\xi$ for $N=30$ and $N=80$. Low values of the index $\xi$ correspond to dominant wavevectors $k$ that start at the Brillouin zone edge and get progressively closer to the value $|k|=k_0$, separating radiative and evanescent fields, as $\xi$ increases.

\tblue{This decay rate suppression of eigenstates of the effective Hamiltonian has important, direct consequences on the full system dynamics, as derived from the system density matrix. In particular, as we show in Sec.\,\ref{sec:liouvillian} and \,\ref{Eigenvalues_Liouvillian}, the density matrix solution of the master equation (\ref{eq:Liouville_master_equation}) can be decomposed in terms of Liouvillian eigenstates $Z_{\Lambda_n}$ with eigenvalue $\Lambda_n.$ These eigenstates evolve under (\ref{eq:Liouville_master_equation}) with a simple time dependence $e^{\Lambda_n t}$. In particular, the spectral gap of the Liouvillian operator -- defined as $\Delta=\min_n {\rm Re}[-\Lambda_n]$, with $\Lambda_n$ the eigenvalues of $\mathcal{L}$ different from zero -- determines the slowest (exponential) relaxation timescale. For our system, one can show that $\Lambda=(\lambda_m-\lambda_n^*)/(i\hbar)$ is an eigenvalue of $\mathcal{L}$ if $\lambda_m$ and $\lambda_n$ are two eigenvalues of $\mathcal{H}_{\rm eff}$\,\cite{Torres14} [see Appendix \ref{Eigenvalues_Liouvillian}]. Therefore, we find that the Liouvillian gap $\Delta=\Gamma^{(1)}_{\xi=1}/2\sim \Gamma_0/N^3$ goes to zero with increasing system size. In 1D arrays, this necessary ingredient for critical slow-down of relaxation dynamics only occurs for lattice spacing $d<\lambda_0/2$, which is difficult to generate with free-space beams. \tbl{However,} for atoms in a 3D lattice (see Sec. \ref{sec:3D}) the closing gap persists up to $d\simeq \lambda_0$, which is easily accessible to experiments.} 

{We now consider multi-excitation eigenstates of $\mathcal{H}_{\rm eff}$.} One numerically finds the existence of multi-excitation subradiant eigenstates, with a cubic suppression of the decay rate with $N$ -- similarly to the single-excitation sector. While single-excitation subradiant states can be interpreted within classical linear optics as guided excitations, the fact that this ``fiber'' is made of non-linear two-level atoms causes multi-excitation subradiant states to have a highly non-trivial character. In particular, the collision of two subradiant single excitations would create a sharp spatial discontinuity in the two-excitation wave function, as a single atom cannot be excited twice. As dissipation occurs in momentum space, the broad momentum distribution associated with this spatial feature induces a large dissipation rate, i.e., the collision of two subradiant excitations can cause them to become unguided. One thus expects the excitations composing a multi-excitation subradiant state to  smoothly repel from each other. One actually finds that low-density multi-excitation subradiant eigenstates are well approximated by anti-symmetric combinations of single-excitation subradiant states (defined by their wave function amplitudes $c_n^k$), thus enforcing {``fermionic'' correlations or Pauli exclusion\cite{asenjo17,wg_paper}}. More specifically, for $m_{\rm ex}=2$ \tbl{excitations}, one finds that an ansatz, $|\psi^{(F)}_{(k1,k2)}\rangle = \mathcal{N}\sum_{m<n}\left[ c^{k_1}_m c^{k_2}_n- c^{k_2}_m c^{k_1}_n\right] \sigma^m_{eg} \sigma^n_{eg} |g\rangle$ -- with $\mathcal{N}$ a normalization factor -- approximates well two-excitation eigenstates of $\mathcal{H}_{\rm eff}$ when $(k_1,k_2)$ are away from $\pm k_0$. Most two-excitation eigenstates can thus be characterized unambiguously by a pair $(k_1,k_2)$ of quantized wavevectors within the first Brillouin zone. As a result of this fermionization, multi-excitation subradiant states contain rich correlations between particles. This can be seen in Fig.\,\ref{Fig1} (a) (iii), where the population of $\ket{e_n,e_m}$ of atoms $n$ and $m$ to be simultaneously excited is plotted for the most subradiant two-excitation eigenstate. One finds an anti-bunching in position, in that the population is peaked {when the excited atoms lie} both far away from each other and from the system boundary. In addition, the decay rate of such states composed of two single-excitation states is found to be comparable to the sum of the single-excitation decay rates $\Gamma^{(2)}_{(k_1,k_2)}\sim \Gamma^{(1)}_{k_1}+\Gamma^{(1)}_{k_2}$. 
This anti-symmetric ansatz and the approximate additivity of multi-excitation decay rates also generalize to higher numbers of excitations.

\tbl{As a mathematical note,} the eigenstates $\ket{\psi_\xi^{(m_{\rm ex})}}$ of the non-Hermitian Hamiltonian $\mathcal{H}_{\rm eff}$ are generally non-orthonormal in the quantum mechanical sense, i.e. $\ex{\psi_\xi^{(m_{\rm ex})}|\psi_{\xi'}^{(m_{\rm ex})}}\neq \delta_{\xi,\xi'}$. \tblue{They however constitute a complete basis, and we have $\mathds{1}=\sum_{m_{\rm ex}}\sum_{\xi} \ket{\psi_{\xi}^{(m_{\rm ex})}}\bra{\varphi_{\xi}^{(m_{\rm ex})}}$, where $\langle \varphi^{(m_{\rm ex})}_{\xi}|$ denotes the left eigenvector of the effective Hamiltonian with the same eigenvalue as $\ket{\psi_{\xi}^{(m_{\rm ex})}}$, i.e. $\langle \varphi^{(m_{\rm ex})}_{\xi}| \mathcal{H}_{\rm eff}=\hbar (\omega_{\xi}^{(m_{\rm ex})}-i\Gamma_{\xi}^{(m_{\rm ex})}/2)\langle \varphi^{(m_{\rm ex})}_{\xi}|$  , with normalization condition $\langle \varphi^{(m_{\rm ex})}_{\xi} |\psi^{(m_{\rm ex}')}_{\xi'} \rangle=\delta_{m_{\rm ex},m_{\rm ex}'} \delta_{\xi,\xi'}$. Aside from these formal statements, it happens to be that for our particular system, the quantum mechanical overlap $\langle \psi^{(m_{\rm ex})}_{\xi}|\psi^{(m_{\rm ex})}_{\xi'}\rangle$ between different eigenstates is found to be small. In particular, the most subradiant eigenstates have been shown to be nearly orthogonal\,\cite{asenjo17}.}

\subsection{1D waveguide model}
In later sections, we present exact numerics for free-space arrays of up to $N=14$ atoms. Here, we present a closely related model, consisting of regularly spaced atoms coupled to an idealized 1D waveguide (Fig.\,\ref{Fig1}\,(b)). This system easily allows for numerical simulations of larger atom number via the matrix product state ansatz, and provides additional evidence for the scaling behavior {seen in free space}. Examining this system is justified because subradiant states in this system exhibit all the same essential properties as in 1D free-space arrays, as we now show.

The dynamics of atoms interacting via photons in a 1D waveguide is governed by the spin model formalism Eq.\,(\ref{eq:Liouville_master_equation}) with\,\cite{Dzsotjan10,Gonzalez-Tudela11,chang12}
\begin{align}
\mathcal{H}_{\rm eff}=&  -i \sum_{m,n=1}^N \frac{\hbar \Gamma_{0}}{2}\exp \left[i k_0  |z_m-z_n| \right]  \sigma^n_{eg} \sigma^m_{ge}\label{eq:Effective_Hamiltonian_wg}.
\end{align}  
Here, $\Gamma_0$ denotes the single atom emission rate into the waveguide, and $k_0$ is the resonant wavevector. The matrix elements for the correlated decay rates, as defined in\,Eq.\,(\ref{eq:Liouville_master_equation}) take on the form $\Gamma_{m,n}= \Gamma_{0}  \cos(k_0 |z_{m}-z_{n}|)$.

We plot in Fig.\,\ref{Fig1} (b) (i) the collective single excitation decay rates for this coupling, indexed by their dominant wavevector $k$, for a fixed number of atoms $N=30$. One finds that radiant eigenstates are localized in reciprocal space around the resonant wavevector $k=\pm k_0$, whereas eigenstates for which $k$ ranges outside these regions are of subradiant nature. Thus, the single-excitation eigenstate \tbl{structures} of the waveguide and free space setup \tbl{show} close similarity in that continuous regions in momentum space are characterized by radiant and subradiant properties. In analogy to the case of an atomic chain in free space, the most subradiant decay rates scale as $\Gamma^{(1)}_{\xi}/\Gamma_{0}\propto \xi^2/N^3$ with $1\leq\xi\leq N$ ordering the eigenstates by increasing decay rates, {as} illustrated in Fig.\,\ref{Fig1}\,(b)\,(ii) for $N=30$ and $N=80$. 
\tbl{Moreover, the most subradiant multi-excitation states can also be constructed out of a fermionic ansatz involving single-excitation states\,\cite{wg_paper}. This is explicitly illustrated in Fig.\,\ref{Fig1}\,(b)\,(iii), for the most subradiant two-excitation state of $N=30$ atoms.}



 \tblue{We demonstrated in this {section} that a 1D lattice clock represents a quantum open critical system in terms of its Liouvillian spectrum, as the slowest exponential timescale associated with relaxation becomes infinite with increasing system size. This peculiar behavior has already been shown to lead to critical slowing down of dynamics of some observables in a number of other open systems \,\cite{Cai13,medvedyeva14,rota17}}. Specific to our system, it raises the following questions: (i) Do dipole-dipole interactions result in an apparent shift of the atomic resonance frequency, as probed by a Ramsey sequence, and how does this shift depend on the interrogation time? (ii) Does subradiance result in a longer excited population and persistence of the clock signal, thus opening the possibility for longer interrogation times? We will begin with the second question, and in particular analyze the time dynamics of the total number of excitations in the system $\hat{n}_e=\sum_n \sigma^n_{ee}$.

\section{\label{sec:decay_dynamics} Population decay : numerics}
In this section, we study numerically the dynamics of a spatially global, highly-excited state, such as states of the form of Eq.\,(\ref{eq:initial_clock_state}) relevant to clocks, or to provide greater generality, a fully inverted state $ \bigotimes_{n=1}^N |e_n \rangle$. Here, we distinguish the case of  ``spatially global'' meaning that locally the atoms have equal excitation probability, to preclude cases where, for example, one section could be highly excited and where excitations could thus propagate or diffuse to other parts of the system. Our analysis suggests that {the system initialized in any such global, highly-excited state has} the following properties {at long times}, which are robust to the specific initial state: (i) the excited population decays in time as a power law, $\sim 1/\sqrt{t}$, and (ii) {fermion-like density-density correlations between excitations emerge due to the persistence of subradiant states}.

\subsection{\label{sec:power_law_observation} Emergence of a power law in the population decay }

\begin{figure}[h!]
\center
\includegraphics[scale=0.6]{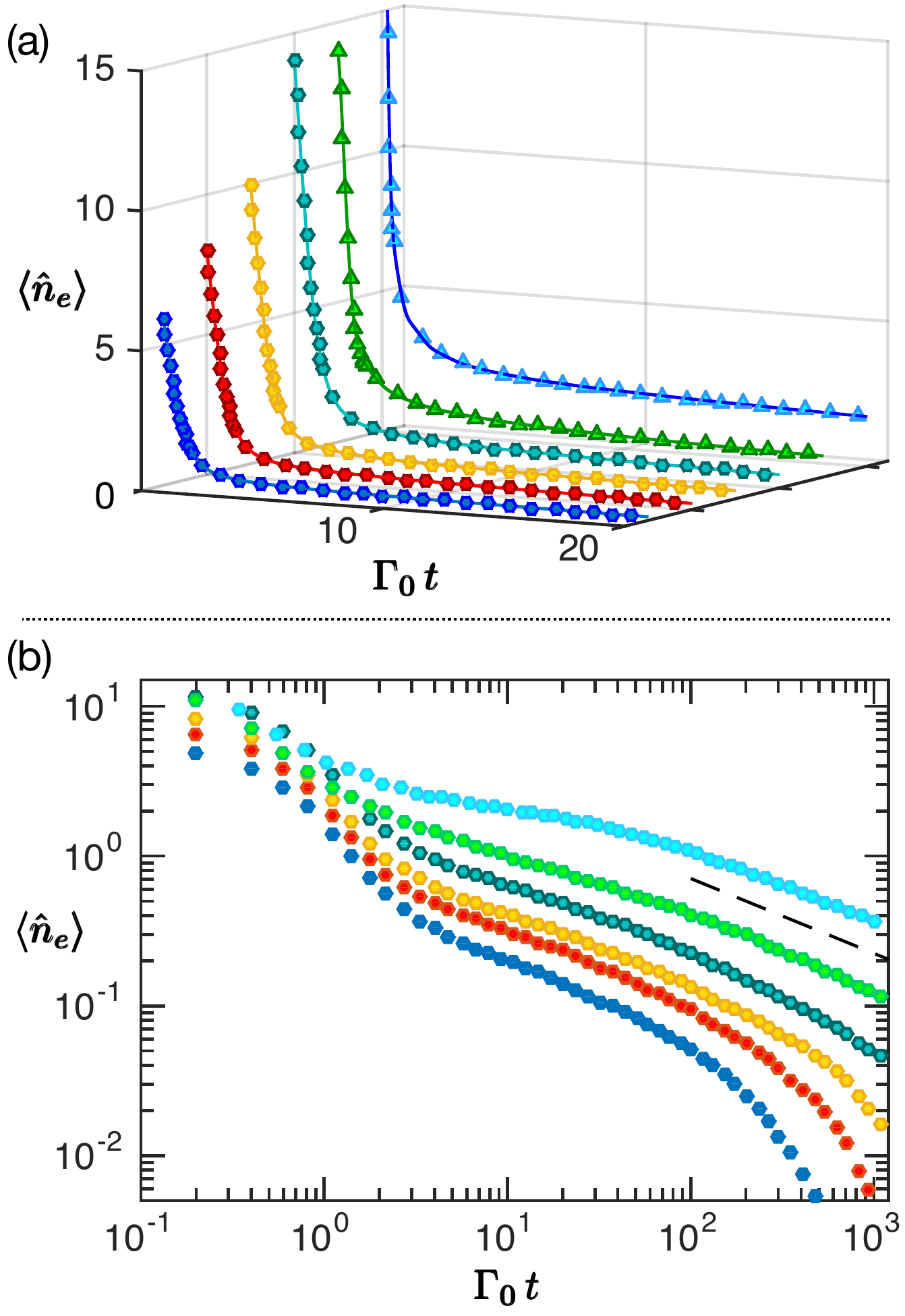}
\caption{ Dynamics of the total population $\langle \hat{n}_e \rangle$ in time {for an initially inverted state,} on (a) a linear scale and (b) a double logarithmic scale. The inter-atomic distance is chosen as $d/\lambda_0=0.1$ (waveguide setup) and $d/\lambda_0=0.2$ (free space setup).  Free space results are depicted by circles for atom numbers $N=6$ (blue), $N=8$ (red), $N=10$ (orange) and $N=14$ (green), and waveguide results by triangles for $N=14$ (green) and $N=30$ (blue). The dashed line in (b) shows a power-law guide to the eye with exponent $\eta=0.5$. 
}
\label{Fig2}
\end{figure}

Fig.\,\ref{Fig2} shows the {numerically obtained} dynamics of the total excited state population $\langle \hat{n}_e \rangle=\langle\sum_n \sigma^n_{ee}\rangle$ for a system initially prepared in the fully inverted state $ \bigotimes_{n=1}^N |e_n \rangle$. Various atom numbers are considered for both a free-space lattice and the waveguide setup. After a fast initial decay at short times (Fig.\,\ref{Fig2}\,(a)), a power-law behavior $\langle \hat{n}_e\rangle(t)\sim (\Gamma_{0}t)^{-\eta}$ emerges at long times (Fig.\,\ref{Fig2}\,(b)),  which becomes more and more pronounced with increasing atom number.  The power-law scaling coefficient is found as $\eta \simeq  0.5$, and a convergence to that value with atom number is found for the waveguide configuration. For a finite-size system, the rate of decay in population eventually returns to exponential as the Liouvillian gap is always finite, as can be seen in Fig.\,\ref{Fig2}\,(b) for the smallest system sizes studied ($N=6$ and $N=8$).  The results shown in Fig.\,\ref{Fig2} were obtained {by equivalently representing the evolution of the master equation\,(\ref{eq:Liouville_master_equation}) via the evolution of a wave function} under stochastic quantum jumps \,\cite{Dalibard92,Dum92,Carmichael93,Plenio98,Daley14} for $N\leq 14$, using an average over $10^4$ trajectories. For the waveguide \tbl{configuration of $N=30$} an MPS simulation \,\cite{schollwoeck11, verstraete04} [see \ref{appendix:MPS}] has been performed. The decay features are found to be robust to the precise choice of the initial state -- provided it is highly excited -- and  the same properties {are observed as well for an initial clock state} (see \ref{sec:moreplaw}).

\begin{figure}[h!]
\center
\includegraphics[scale=0.55]{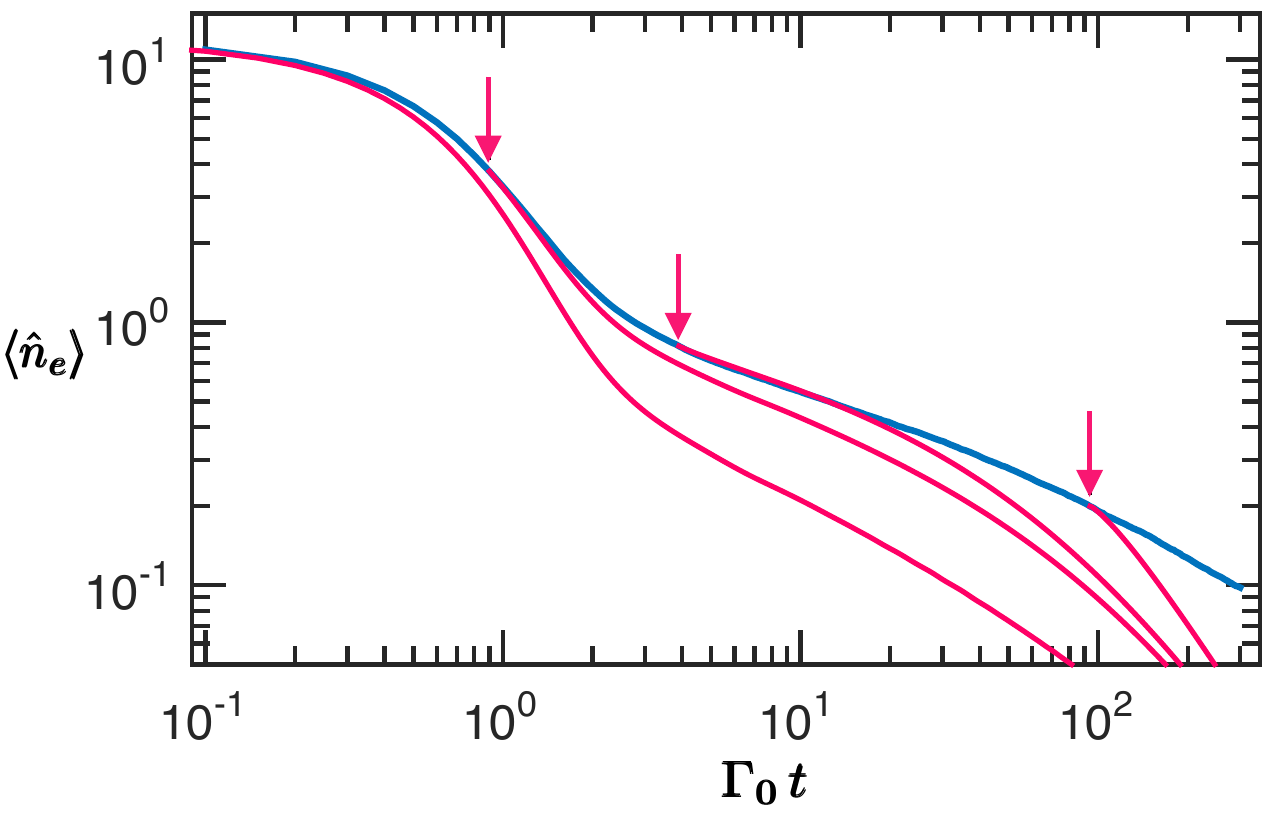}\\
\caption{Excited state population decay for an initially fully excited state and a free-space chain of $N=12$ atoms and $d=0.2\lambda_0$. The blue solid line corresponds to solving the spin-model master equation (in the wavefunction quantum jump framework), red lines to a second order mean field calculation. The mean field calculation has been performed for different initial times ($t = 0$ and the times indicated by arrows), where the corresponding initial conditions are obtained from the master equation results.
}
\label{Fig_meanfield}
\end{figure}

In Fig.\,\ref{Fig_meanfield}, we compare the full population dynamics to a mean field-like approach. In particular, we numerically solve the equations of motion for the expectation values of the populations, $\langle \dot{\sigma}_{ee}^{i}\rangle=\textrm{Tr}\left(\dot{\rho} \sigma_{ee}^{i}\right)$. Under $\mathcal{L}[\rho]$, the dynamics of n-body operators generally depend on ($n+1$)-body operators. Here, we truncate the correlations to two-body, by approximating $\langle A_i B_j C_k \rangle=\langle A_i B_j \rangle \langle C_k \rangle+\langle A_i\rangle\langle B_j C_k \rangle+\langle A_i C_k \rangle\langle B_j \rangle-2\langle A_i \rangle\langle B_j\rangle\langle C_k \rangle$, where $A$, $B$ and $C$ are local one-body operators at distinct positions $i$, $j$ and $k$ \,\cite{kramer15}. We can also evolve the resulting equations not only from $t=0$, but starting from an arbitrary initial time $t_i>0$ (indicated by arrows in Fig.\,\ref{Fig_meanfield}), using as initial conditions the numerically exact correlation functions at $t_i$ obtained by full simulations. Interestingly, the mean field dynamics seem to diverge from the full solution regardless of initial time $t_i$, with the former predicting a more rapid decay of population. This suggests that at each stage of the evolution, highly correlated states, {such} as the eigenstates discussed previously, play a crucial role. 


Such an algebraic decay in reaching a steady state has already been predicted for specific many-body systems described by a Liouville master equation \cite{Cai13,medvedyeva14,rota17}, and for which the spectral gap closes in the thermodynamic limit. A first argument to justify the emergence of such a power law behavior consists of analyzing the density of states with decay rate close to zero in the thermodynamic limit \cite{medvedyeva14}. As an illustrative example, assuming a set of long-lived states $\xi$ contributes equally, one estimates the population in the long-time limit to be $\langle \hat{n}_e \rangle (t)  \sim \sum_{\xi} e^{-\Gamma_{\xi}t}$. The {smooth distribution of subradiant decay rates approaching zero for large $N$ then allows for a continuous description of the long-time evolution of the population}. This results in $\langle \hat{n}_e  \rangle(t) \sim \int_0^{\infty} {\rm d}\Gamma \mathcal{D}(\Gamma) e^{-\Gamma t}$, where $\mathcal{D}(\Gamma)$ is the density of states with decay rate $\Gamma$. As a first assumption, we consider just the single-excitation subradiant states. Their scaling $\Gamma_{\xi}\propto \Gamma_{0} \xi^2/N^3$, identified in the previous section for the most subradiant states for both the waveguide and the free-space setup, leads to a density of states $\mathcal{D}(\Gamma)\sim \Gamma^{\kappa}$ at small $\Gamma$, with $\kappa\sim -0.5$. Evaluating the integral leads to $\langle \hat{n}_e \rangle(t)\sim 1/(\Gamma_{0}t)^{1+\kappa}\sim (\Gamma_0 t)^{-1/2}$, in approximate agreement with the numerical calculations.

We point out that this naive argument is not complete, as it wrongly implies that one has to wait until the contribution of multi-excitation states vanishes. Moreover, it ignores how such states decay into lower-excitation states. A more rigorous {argument based upon the diagonalization of the Liouvillian} is provided in Sec.\,\ref{sec:power_law_from_Liouvillian}.



\subsection{\label{sec:convergence_2_exc} \tbl{Buildup of fermionic density-density correlations}}
\begin{figure}[h!]
\center
\includegraphics[scale=0.5]{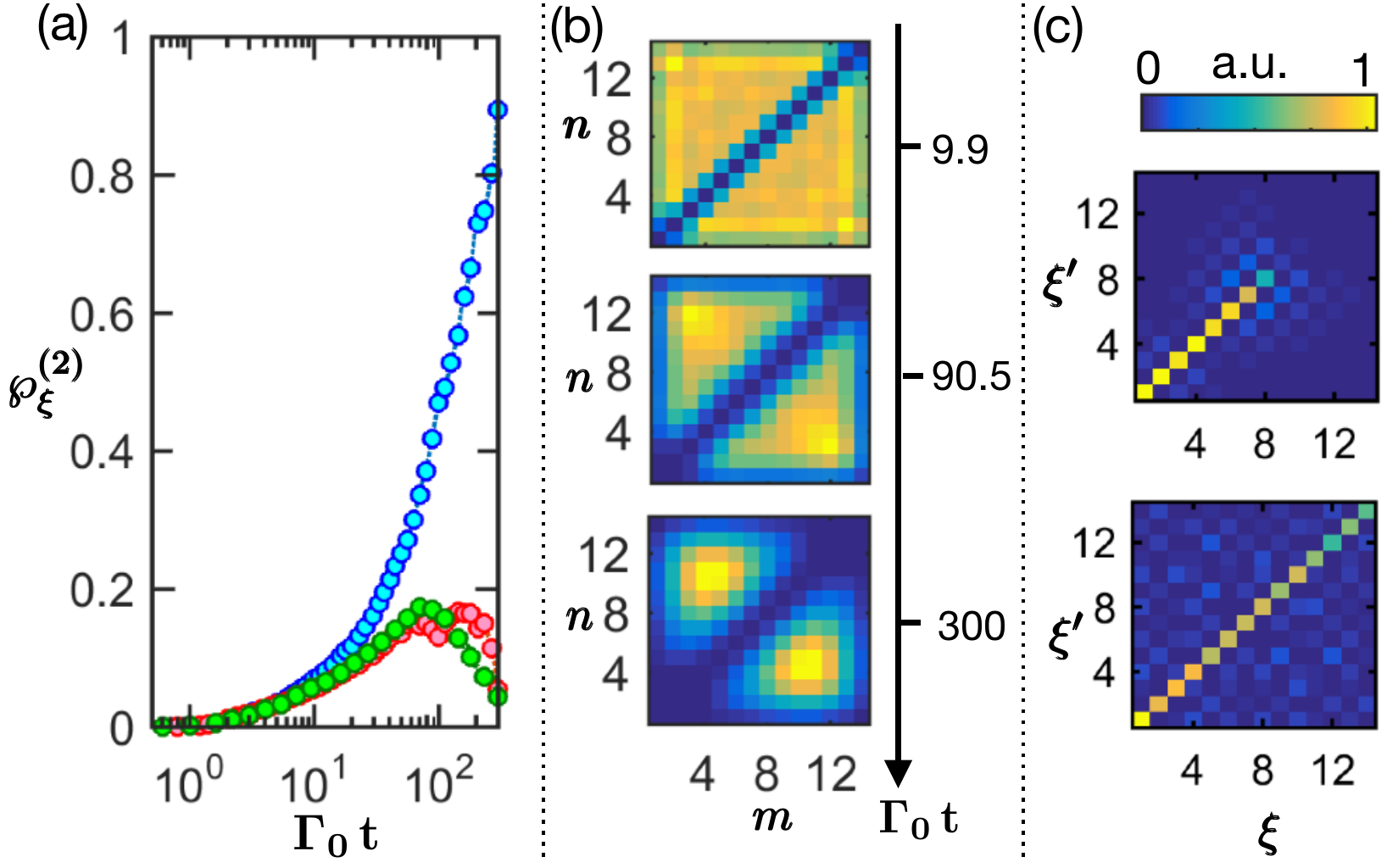}\\
\caption{(a) Contribution of the three eigenstates of lowest decay ($\xi = 1$ (blue), $2$ (red) and $3$ (green)) to the population in the two-excitation subspace, for the decay dynamics as in Fig.\,\ref{Fig2} and for the free space configuration with $N=14$ and $d/\lambda_0 = 0.2$. (b) Two-excitation correlations $\ex{\sigma_{ee}^{n}\sigma_{ee}^{m}}$ at selected times. (c) Population and coherences $|\ex{\psi_{\xi'}^{(m_{\rm ex})}|\rho(t)|\psi_{\xi}^{(m_{\rm ex})}}|$ for $t=11\,\Gamma_{0}^{-1}$ and the single $m_{\rm ex}=1$ (upper) and two-excitation $m_{\rm ex}=2$ (lower) manifold.}
\label{Fig_convergence}
\end{figure}

Both radiant and subradiant eigenmodes are involved in the decay process, reflected by the fast initial decay and the persistence of population at long timescales. One can therefore expect that observables acquire the properties of the most subradiant states in the long-time limit. To provide an example at the microscopic level, we can consider the projection of the entire system density matrix $\rho(t)$ into the two-excitation subspace, $\rho^{(2)}(t)$ (here, we renormalize so that $\textrm{Tr}(\rho^{(2)}(t))=1$). In Fig.\,\ref{Fig_convergence}\,(a), we plot the overlap $p_\xi^{(2)}=\ex{\psi_\xi^{(2)}|\rho^{(2)}(t)|\psi_\xi^{(2)}}$ of that state in time $t$ with the three most subradiant two-excitation eigenstates ($\xi=1,2,3$) for an atomic chain of $N=14$ atoms. One finds a convergence to the most subradiant eigenstate $\xi=1$, which becomes the dominant contribution for $\Gamma_0\,t \gtrsim 20$. 
The microscopic picture provided above also manifests itself in macroscopic observables. For example, in Fig.\,\ref{Fig_convergence}\,(b) we plot the density-density correlations $\ex{\sigma_{ee}^m\sigma_{ee}^n}$ of excitations (now calculated over the entire system density matrix $\rho(t)$). These correlations are plotted for various specific times $t$ during the evolution. It can be seen that these correlations exhibit increasingly ``fermionic'' character in time, and at sufficiently long times essentially reflect that of the most subradiant two-excitation state.

Interestingly, such a convergence to the most subradiant eigenstates in low-lying excitation manifolds can be shown to be of universal nature --- with the only requirement that the system is initially prepared in a highly excited state. A decay from such a highly excited state populates all low-lying eigenstates in a smooth way, a property we elaborate on at a later stage, such that subradiant states become significantly populated and finally dominant at long times due to their largely inhibited decay. Apart from the details of the convergence process and the final populations, the convergence is independent of both the specific form of the initial state and the specific atom chain configuration. 

\section{Population decay : a jump picture}\label{sec_jumppic}
A significantly simplified model of the decay dynamics can be obtained by only taking eigenstate populations into account. The validity of such an approximation is motivated in Fig.\,\ref{Fig_convergence}\,(c), where the density matrix elements in the single- [$m_{\rm ex}=1$] and two-excitation [$m_{\rm ex}=2$] manifolds are plotted at a fixed time $t=11\Gamma_0^{-1}$ after the preparation of a fully excited state. Specifically, density matrix  elements are depicted in the eigenstate basis $\rho_{\xi',\xi}^{(m_{\rm ex})}\equiv\ex{\psi_{\xi'}^{(m_{\rm ex})}|\rho(t)|\psi_\xi^{(m_{\rm ex})}}$, where elements $\xi=\xi'$ and $\xi\neq\xi'$ are denoted as populations and coherences, respectively. From Fig.\,\ref{Fig_convergence}\,(c) it follows that only populations are {significant}; the weak coherence contributions can be shown to originate predominantly from the non-orthogonality of the eigenstate basis as discussed in Sec.\,\ref{subsec:subradiant_properties}. More generally, one numerically finds that only populations contribute in the long-time limit and in particular that decay processes only destroy but do not generate coherences.

Based on that insight, we now analyze the decay dynamics, or more specifically the dynamics of eigenstate populations, based on a rate model. In that model, coherences are neglected, and populations of eigenstates are connected by transition rates. While the eigenstates do not formally constitute an orthogonal basis, we find empirically that this rate model agrees very well with the full numerics. It should be noted that as the eigenstates themselves and their decay properties arise from many-body interactions, the agreement of a rate model does not imply that the system is nearly classical (cf. Fig.\,\ref{Fig_meanfield}, where we showed that correlations play an important role in the dynamics).

\subsection{Transition rate between eigenstates}\label{sec_jeig}

The transition rate $\gamma_{\xi,\xi'}^{(m_{\rm ex})}$ from state $\xi$ in the excitation manifold $m_{\rm ex}$ to state $\xi'$ in manifold $m_{\rm ex}-1$ can be defined as
\begin{align}\label{trans_rate1} \gamma_{\xi,\xi'}^{(m_{\rm ex})} =& \textrm{Tr}\bigl(\ket{\psi_{\xi'}^{(m_{\rm ex}-1)}}\bra{\psi_{\xi'}^{(m_{\rm ex}-1)}}\,\,\mathcal{J}[\ket{\psi_\xi^{(m_{\rm ex})}}\bra{\psi_\xi^{(m_{\rm ex})}}] \bigr) \notag\\
&\times\mathcal{N}_\xi^{(m_{\rm ex})}
\end{align}
where $\mathcal{J}[\rho]=\sum_{m,n}\Gamma_{m,n}\sigma_{ge}^m\rho\sigma_{eg}^{n}$ is the decay contribution of the Liouvillian as defined in Sect.\,\ref{sec_spinmodel}. Note that $\Gamma_{\xi}^{(m_{\rm ex})}=\textrm{Tr}\bigl(\mathcal{J}[\ket{\psi_\xi^{(m_{\rm ex})}}\bra{\psi_\xi^{(m_{\rm ex})}}]\bigr)$ corresponds to the total decay rate of state $\ket{\psi_{\xi}^{(m_{\rm ex})}}$. Rates to individual states as defined in Eq.\,(\ref{trans_rate1}) are obtained by an additional projection onto these states $\ket{\psi_{\xi'}^{(m_{\rm ex}-1)}}$. As eigenstates here are non-orthonormal, we enforce a normalization $\mathcal{N}_\xi^{(m_{\rm ex})}$ such that the total decay rate is preserved $\sum_{\xi'}\gamma_{\xi,\xi'}^{(m_{\rm ex})}=\Gamma_{\xi}^{(m_{\rm ex})}$. \tblue{Generally, for sufficiently many atoms, the eigenstates are almost orthonormal and transition rates are well-approximated even without this
additional normalization step, i.e. by setting $\mathcal{N}_{\xi}^{(m_{\mathrm{ex}})}\simeq 1$.}

\subsection{Decay structure of subradiant eigenstates}\label{sec_dlow} 
\begin{figure}[h!]
\center
\includegraphics[scale=0.55]{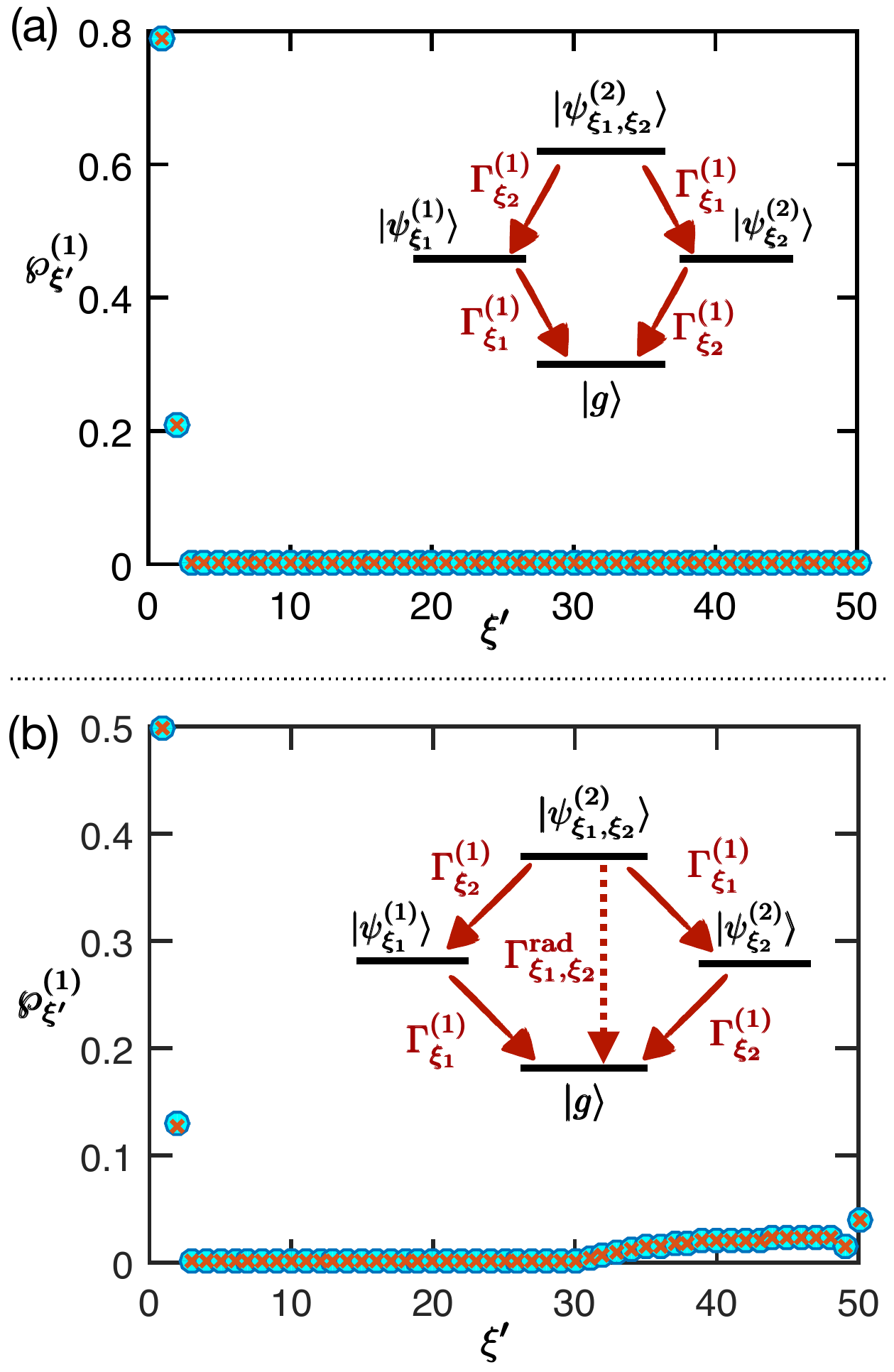}\\
\caption{Probability $\wp_{\xi'}^{(1)}$ for the most subradiant two-excitation eigenstate to decay into the single-excitation eigenstates $\xi'$ for $N=50$ and in (a) the waveguide ($d=0.1\lambda_0$) and (b) the free space ($d=0.2\lambda_0$) setup. Here, $\xi'$ indexes the eigenstates by increasing decay rates. Blue circles are obtained from the transition rate model of Sect.\,\ref{sec_jumppic} {and} red crosses from the Liouvillian eigenstate expansion of Sect.\,\ref{eigenstate_construction}.  The inset depicts the decay channels and rates, valid up to $1/N^2$ corrections in the atom number $N$. {In both free space and the waveguide, the state has substantial probability to decay into the two single-excitation states ($\xi'=1,2$) from which it is composed, as illustrated in the insets. In free space, the state also has a non-negligible probability to decay into superradiant ($\xi'\gtrsim 30$) states. This results in an effective two-excitation loss to the ground state (dashed arrow in the inset),} at a rate $\Gamma_{\xi_1,\xi_2}^{\rm rad}=\Gamma_{\xi_1,\xi_2}^{(2)}-\Gamma_{\xi_1}^{(1)}-\Gamma_{\xi_2}^{(1)}$.}
\label{fig:structure}
\end{figure}

Based on the transition rates defined above, we now analyze the decay structure of two-excitation eigenstates. Starting in such a state $\ket{\psi_\xi^{(2)}}$, the emission of a single photon transfers the system to the single-excitation manifold. A specific eigenstate $\ket{\psi_{\xi'}^{(1)}}$ in that manifold is reached with probability $\wp_{\xi'}^{(1)}=\gamma_{\xi,\xi'}^{(2)}/\Gamma_\xi^{(2)}$. 
That probability -- for the initial state being the most subradiant two-excitation eigenstate $\ket{\psi_{\xi=1}^{(2)}}$ -- is illustrated in Fig.\,\ref{fig:structure} for both the waveguide and free-space setup.

 For the waveguide configuration, the decay overwhelmingly populates the two most subradiant single excitation states $|\psi^{(1)}_{\xi=1}\rangle$ and $|\psi^{(1)}_{\xi=2}\rangle$, with rates $\gamma_{1,1}^{(2)}\simeq\Gamma^{(1)}_{\xi=2}$ and $\gamma_{1,2}^{(2)}\simeq\Gamma^{(1)}_{\xi=1}$, respectively. In other words, the two-excitation state is given approximately by an anti-symmetric combination of two single-excitation states, and these constituent states decay approximately independently [see inset of Fig\,\ref{fig:structure}\,(a)]. As one consequence, the two-excitation state is more likely to decay into the \textit{more} subradiant single-excitation state from which it is composed. More generally, one finds that any subradiant two-excitation state decays into the two single-excitation states it is composed of. The relative weight of other decay channels vanishes with atom number $N$ as $1/N^2$. 
Such a decay structure extends to all of the highly subradiant states in the low excitation manifolds ($m_{\rm ex}\ll N$), which implies that their decay can be interpreted as the gradual decay of their single-excitation constituents.
 This also offers a simple interpretation of the decay rate addition in subradiant manifolds \cite{wg_paper}, i.e. the fact that $\Gamma^{(m_{\rm ex})}_{(\xi_1,...,\xi_{m_{\rm ex}})}\simeq \Gamma^{(1)}_{\xi_1}+...+\Gamma^{(1)}_{\xi_{m_{\rm ex}}}$.

In free space, we observe a qualitatively similar decay structure with an additional non-zero probability to decay into superradiant states [see Fig.\,\ref{fig:structure}\,(b)]. Compared to the subradiant evolution timescale, these latter states decay almost instantaneously, and therefore this additional decay channel can be seen as an effective `direct' decay from the doubly excited to the ground state [see the dotted red arrow in the schematic of Fig.\,\ref{fig:structure}\,(b)]. This observation of an additional channel is in line with a moderately enhanced decay rate of two-excitation states as compared to the sum of their single-excitation component rates, i.e. $\Gamma^{(2)}_{(\xi_1,\xi_2)}/ (\Gamma^{(1)}_{\xi_1}+\Gamma^{(1)}_{\xi_2})\sim 1+u$ with $u\simeq 0.6$. An {analogous} decay structure can be found for subradiant states of higher excitations. However, the relative importance of the decay channel via superradiant states, over the ``shedding'' of constituent single-excitation states, increases with the number $m_{\rm ex}$ of excitations. Specifically, we find a fraction $\sim (m_{\rm ex}-1) u/[1+(m_{\rm ex}-1)u]$  of the total decay rate directed towards superradiant states. Physically, we attribute this enhancement to the fact that collisions between excitations can cause radiation loss from the bulk of the array, and not only from the ends.

\subsection{Excitation-hole symmetry}\label{sec_jexhole}
As seen in Sect.\,\ref{Sect2}, the properties of eigenstates in the low-excitation manifolds are well-understood. In particular, {single-excitation states} can be associated with a wavevector $k$, which crucially determines the decay properties. Subradiant states of several but few excitations can be composed {from anti-symmetric combinations} of single-excitation eigenstates, and thus inherit the properties of their constituents. In particular, we have seen in the previous section how this composition explains their decay behavior. 

Here, we show that highly excited states can be viewed in terms of the small number of ``holes'' corresponding to atoms in their ground states. A symmetry between highly excited states with holes and few-excitation states allows one to identify the salient properties of the former. 

The excitation-hole symmetry follows directly from the effective Hamiltonians\,(\ref{eq:Effective_Hamiltonian}) and (\ref{eq:Effective_Hamiltonian_wg}). Separating this Hamiltonian into diagonal $\mathcal{H}_{\rm ind}={-i\hbar (\Gamma_0/2)}\,\sum_n \sigma_{ee}^n$ and non-diagonal exchange contributions $\mathcal{H}_{\rm ec}=-\mu_0\omega_0^2 \sum_{m\neq n} {\bf{p}}^\dagger\,\doublearrow{{\bf{G}}}({\bf{r}}_n,{\bf{r}}_m,\omega_0)\,{\bf{p}}  \sigma^n_{eg} \sigma^m_{ge}$, one straightforwardly finds the exchange contribution to remain invariant under an excitation-hole exchange $\ket{g}\leftrightarrow \ket{e}$. Such an invariance does not hold true for the diagonal term; however, that one is constant within an excitation manifold. Therefore, applying an excitation-hole exchange on an eigenstate produces another eigenstate. This implies that eigenstates of $(N-m_{\rm ex})$ excitations can be constructed out of eigenstates of $m_{\rm ex}$ excitations by such an exchange
\begin{equation}\label{exhole1}  \ket{\psi_\xi^{(N-m_{\rm ex})}} = \ket{\psi_\xi^{(m_{\rm ex})}}\,\bigr|_{g\leftrightarrow e} \end{equation}
{with equal energy shifts and a decay rate just differing} by an excitation-manifold dependent contribution arising from $\mathcal{H}_{\rm ind}$,
\begin{align}
&{\omega_\xi^{(N-m_{\rm ex})}=\omega_\xi^{(m_{\rm ex})}\,,}\\
&\Gamma_\xi^{(N-m_{\rm ex})}=\Gamma_\xi^{(m_{\rm ex})}+ (N-2\,m_{\rm ex})\Gamma_0\,.\label{exhole2}
\end{align}
As an example, states of $N-1$ excitations take on the form  $\ket{\psi_k^{(N-1)}}=\sum_n e^{i k d n} \sigma_{\rm ge}^n\,\ket{e}^{\otimes N}/\sqrt{N}$ in the infinite chain limit, and thus represent hole excitations of wavevector $k$ with respect to the totally excited state. From Eq.\,(\ref{exhole2}) it follows that wavevectors $k$ which decay fastest (slowest) in the single-excitation manifold also decay fastest (slowest) in the $N-1$ excitation manifold, albeit the manifold contribution $(N-2m_{\rm ex})\Gamma_0$ makes the latter ones almost equal. 

An equivalence between excitation-hole exchanged state pairs can also be identified for the transition rate Eq.\,(\ref{trans_rate1}).  
The transition rate, for sufficiently many atoms such that eigenstates are nearly orthonormal, can be approximated by $\gamma_{\xi,\xi'}^{(m_{\rm ex})}\simeq \sum_{n,o}(\Gamma_{n,o}/2)\ex{\psi_{\xi'}^{(m_{\rm ex}-1)}|\sigma_{ge}^n|\psi_{\xi}^{(m_{\rm ex})}}\ex{\psi_{\xi}^{(m_{\rm ex})}|\sigma_{eg}^o|\psi_{\xi'}^{(m_{\rm ex}-1)}}$, which by an excitation-hole exchange and using Eq.\,(\ref{exhole1}) and $\Gamma_{n,o}=\Gamma_{o,n}$ leads to 
\begin{equation} \gamma_{\xi',\xi}^{(N+1-m_{\rm ex})}\simeq \gamma_{\xi,\xi'}^{(m_{\rm ex})}\,.  \end{equation} 
Therefore, transition rates between a pair of eigenstates and its excitation-hole inverted analogue are equal. 

\begin{figure*}[!htb]
\center
\includegraphics[scale=0.4]{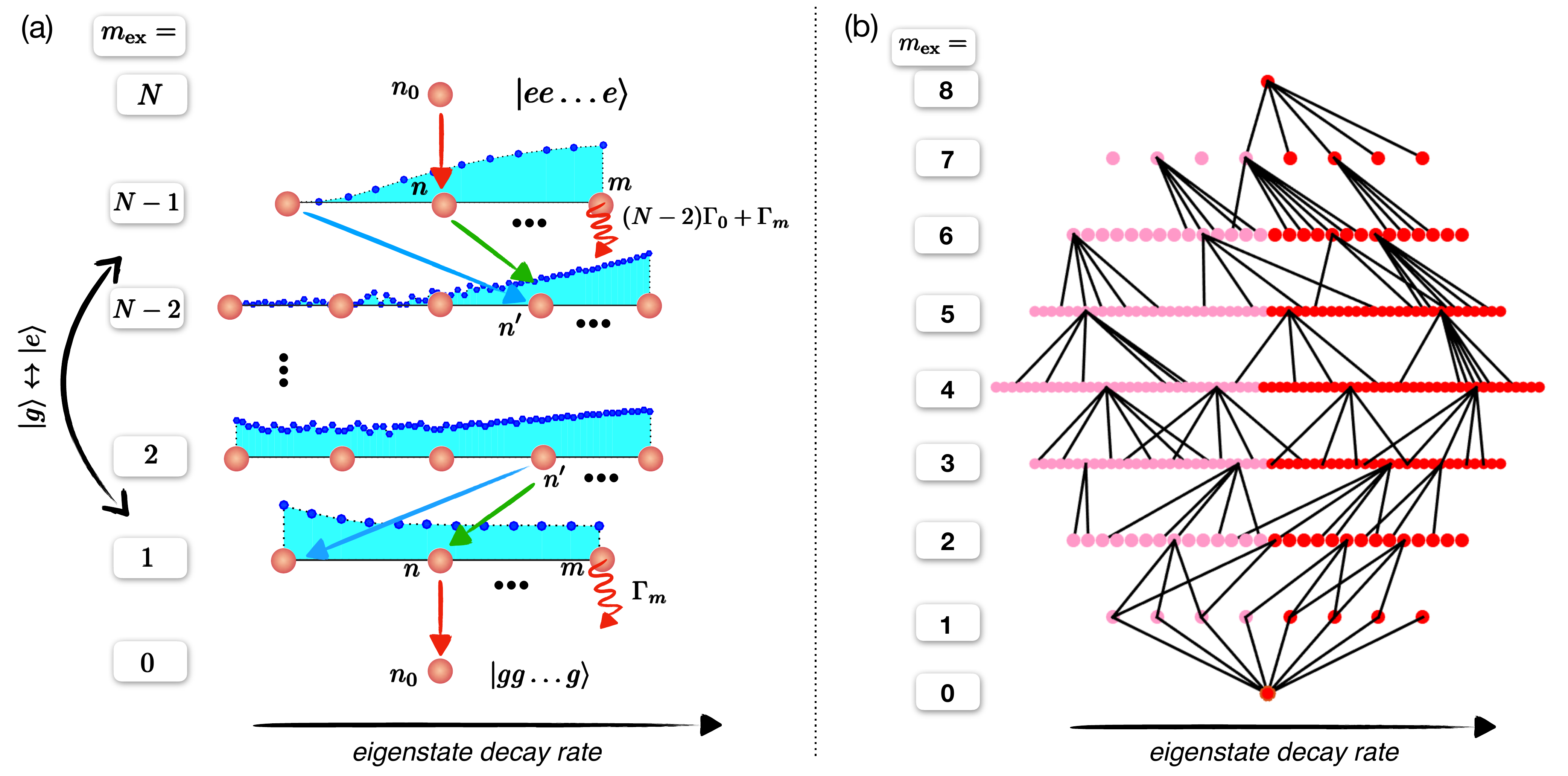}
\caption{\label{Fig3} (a) Schematic of the excitation hole symmetry. States with the same label (e.g., $n_0$, $n$, $n'$) in the upper and lower excitation manifolds are connected by an excitation-hole exchange. Transitions have equal rates when denoted by arrows of the same color. The blue distribution corresponds to the probability for passing through the individual eigenstates (blue markers) in the decay process, for an initial excitation in $\ket{e}^{\otimes N}$ and a free-space atomic chain configuration with $N=12$ and $d=0.4\,\lambda_0$. {In particular, it should be noted that the distribution becomes nearly flat for low numbers of excitation.} (b) Most likely transitions for selected eigenstates in an atomic free-space chain of $N=8$  and $d=0.4\lambda_0$. Eigenstates with $\Gamma_\xi^{(m_{\rm ex})}\geq m_{\rm ex}\Gamma_0$ ($\Gamma_\xi^{(m_{\rm ex})}< m_{\rm ex}\Gamma_0$) are represented by red (pink) circles. For selected eigenstates $\xi$ in the $m_{\rm ex}$-excitation manifold, we calculate the decay rates $\gamma_{\xi,\xi'}$ to eigenstates $\xi'$ in the $m_{\rm ex}-1$ manifold, and indicate by black connecting lines the most likely decay channels. For the figure, these are defined to be all transitions with rates $\gamma_{\xi,\xi'} > 0.5 \,\,{\rm max}_{\xi'} \gamma_{\xi,\xi'}$.
}
\end{figure*}

A schematic of the excitation-hole analogy is depicted in Fig.\,\ref{Fig3}\,(a).  In particular, we visualize all of the eigenstates on a two-dimensional axis, with the vertical axis denoting the number of excitations (with $0\leq m_{\rm ex} \leq N$). The horizontal axis orders the decay rate, with smallest to largest going from left to right (the positions are qualitative, in that two states in different number manifolds with the same horizontal position do not necessarily have the same decay rate). Several selected few-excitation eigenstates are labeled ($n'$,$n$, and $n_0$, with $n_0$ denoting the ground state), and their excitation-hole inverted counterparts are labeled by the same letters. Moreover, transitions of equal rates are indicated by arrows of equal colors. 

As one consequence of the excitation-hole symmetry of transition rates, just as a few-excitation state tends to decay towards a more subradiant state, a few-hole state tends to decay toward a more radiant state. That behavior is illustrated in Fig.\,\ref{Fig3}\,(b), where the most likely decay paths {are depicted by black lines} for selected eigenstates. In high (low) excitation manifolds the paths are directed more towards radiant (subradiant) states.

\subsection{Decay of a highly excited state}
We now utilize the rate-model picture to gain insight into the decay structure of an initially highly excited state. That is, we analyze the decay through the different excitation manifolds and their corresponding eigenstates. The (time-independent) probability $\wp_{\xi}^{(m_{\rm ex})}$ for passing through an eigenstate $\ket{\psi_{\xi}^{(m_{\rm ex})}}$ at some point during the decay process can be iteratively defined as
\begin{equation}\label{eq_rmodel2} \wp_{\xi'}^{(m_{\rm ex}-1)} = \sum_{\xi}\wp_\xi^{(m_{\rm ex})}\,\gamma_{\xi,\xi'}^{(m_{\rm ex})}/\Gamma_{\xi}^{(m_{\rm ex})}.   \end{equation}  
This quantity is given by {the sum of} eigenstate probabilities in the higher manifold (here: $m_{\rm ex}$) multiplied by the probabilities for these states to decay into the state of interest (here: $\ket{\psi_{\xi'}^{(m_{\rm ex}-1)}}$). Clearly, if the initial state has an excitation number greater than $m_{\rm ex}$, and given that in our (finite) system of interest there are no excited eigenstates with zero decay rate, the sum of probabilities of passing through any state within the manifold is $\sum_{\xi} \wp_{\xi}^{(m_{\rm ex})}=1$. 

The probability is illustrated -- for selected excitation manifolds $m_{\rm ex}$ and for an initially fully excited state -- in Fig.\,\ref{Fig3}\,(a) for $N=12$ atoms (see blue markers). In high-excitation manifolds, mostly eigenstates of large decay rate are populated, i.e. {the} probability distribution is strongly weighted towards the right in the figure (for the eigenstates sorted by increasing decay rate from left to right). In low-excitation manifolds the probability distribution of eigenstates is observed to become rather flat. This implies that a redistribution of populations towards more subradiant states takes place in the low-excitation sector. The interpretation of that peculiar decay behavior directly follows from the conclusions of Sect.\,\ref{sec_dlow} and \ref{sec_jexhole}, namely that eigenstates of high (low) excitation numbers tend to  decay towards more (less) radiant states.  Note that the redistribution and the resulting quasi-equal probabilities of transitioning through any given state is one of the crucial mechanisms for the observation of a power-law decay behavior.

Apart from considering probabilities, the rate model also enables to solve for the time-dependent populations $p_{\xi}^{(m_{\rm ex})}(t)$ of each eigenstate during the decay process. The rate equation in that case is given by
\begin{equation}\label{eq_rmodel1} \frac{\mathrm{d}}{\mathrm{d}t}p_{\xi'}^{(m_{\rm ex})}(t) = -\Gamma_{\xi'}^{(m_{\rm ex})}\,p_{\xi'}^{(m_{\rm ex})}(t)+\sum_\xi \gamma_{\xi,\xi'}^{(m_{\rm ex}+1)} p_{\xi}^{(m_{\rm ex}+1)}(t)  \end{equation}
where the first contribution on the right-hand side represents the population loss (decay) of the specific eigenstate $\ket{\psi_{\xi'}^{(m_{\rm ex})}}$, and the second term the population gain through transitions from the excitation manifold above (from states $\{ \ket{\psi_{\xi}^{(m_{\rm ex}+1)}} \}$). That set of equations can be iteratively solved, starting from the uppermost excited state and the initial populations. For an initially fully inverted state, we find good agreement of the population dynamics between the (exact) master equation and the rate model approach (see\,\ref{append_rmod}).

To conclude this section, we observe that the salient properties of decay of few-excitation subradiant states can be intuitively understood by the existence of well defined decay channels, that highly excited states can be equally understood via an excitation-hole symmetry, and that a rate equation works well to describe the population dynamics. {This provides the first comprehensive understanding of collective decay properties in atomic arrays beyond the single-excitation manifold.} While in these dynamics we keep explicit track of the entire large number of discrete eigenstates, an interesting question for future investigation would be whether one can generate an effective ``continuum'' model, which would then allow quantitative predictions for much larger atom number where tracking individual eigenstate populations becomes unfeasible.

\section{Population decay : Liouvillian eigenstructure \label{sec:liouvillian}}

In the previous section, we described the population dynamics with a semi-classical rate equation. Here, we justify the use of such an approach by examining the eigenstructure of the Liouvillian in the low-excitation sectors. We confirm in particular the emergence of an effective single-particle picture for the decay dynamics within subradiant manifolds. Decomposing the initial density matrix in terms of Liouvillian eigenstates finally provides an additional interpretation of the power-law behavior observed at long times in the dynamics of the population.


\subsection{ \label{eigenstate_construction} Liouvillian eigenstates in low-excitation sectors}
{We recall that the master equation\,(\ref{eq:Liouville_master_equation}) describing atomic dynamics under dipole-dipole interactions, $\dot{\rho}=\mathcal{L}[\rho]$, can be decomposed into the form $\mathcal{L}=\mathcal{K}+\mathcal{J}$. $\mathcal{K}$ and $\mathcal{J}$ represent the coherent-like part of the evolution and the jump part, respectively.} The ground state density matrix, $\rho^{(0)}=|g\rangle \langle g |$, is an eigenstate of the Liouvillian $\mathcal{L}$ with eigenvalue 0 (here and in the following, $|g\rangle$ {will denote} the many-body ground state $|g\rangle^{\otimes N}$). For a finite system, $\rho^{(0)}$ is the unique possible steady state of the dynamics. An instructive way to build other eigenstates of $\mathcal{L}$ consists in examining the dynamics of the system when initially prepared in a low-excitation eigenstate of the effective Hamiltonian.  Let us first consider the dynamics of an initial density matrix of the form $\rho^{(1)}_{\xi}=|\psi^{(1)}_{\xi}\rangle \langle \psi^{(1)}_{\xi}|$, where $|\psi^{(1)}_{\xi}\rangle$ is an eigenstate of $\mathcal{H}_{\rm eff}$ in the single-excitation sector. The initial density matrix $\rho^{(1)}_{\xi}$ is an eigenstate of $\mathcal{K}$, with $\mathcal{K}[\rho^{(1)}_{\xi}]=-\Gamma^{(1)}_{\xi} \rho^{(1)}_{\xi}$. We find furthermore that the term $\mathcal{J}$ accounting for jumps between different excitation manifolds gives $\mathcal{J}[\rho^{(1)}_{\xi}]=\Gamma^{(1)}_{\xi} |g\rangle \langle g|$, i.e. $\mathcal{J}$ brings a pure single-excitation mode to the ground state. The traceless operator $Z=\rho^{(1)}_{\xi}-|g\rangle \langle g|$ therefore constitutes an eigenstate of the Liouvillian $\mathcal{L}$ with eigenvalue $-\Gamma^{(1)}_{\xi}$. Decomposing $\rho_\xi^{(1)}$ in terms of $\rho^{(0)}$ and $Z$, we simply recover that
\begin{align}
\rho (t)= e^{-\Gamma^{(1)}_{\xi}t} \rho^{(1)}_{\xi} +(1-e^{-\Gamma^{(1)}_{\xi}t}) |g\rangle \langle g|,
\end{align}
describing the decay of a single-excitation state toward the many-body ground state of the system.

One can extend the construction above for a system initially starting in the pure state $\rho^{(2)}_{\xi}=|\psi^{(2)}_{\xi}\rangle \langle \psi^{(2)}_{\xi}|$ in the two-excitation sector. This initial density matrix is an eigenstate of $\mathcal{K}$, with eigenvalue $-\Gamma^{(2)}_{\xi}$. The action of $\mathcal{J}$, i.e. the loss of one excitation, brings $\rho^{(2)}_{\xi}$ in a superposition of elements of the form $|\psi^{(1)}_{\eta}\rangle \langle \psi^{(1)}_{\mu}|$. A subsequent loss of one excitation brings these latter elements to the ground state $|g\rangle \langle g|$. One can therefore construct an eigenstate $Z$ of $\mathcal{L}$ with eigenvalue $-\Gamma^{(2)}_{\xi}$ under the form,
\begin{align}
Z= \rho^{(2)}_{\xi}+\sum_{(\eta,\mu)=1}^N\alpha_{\eta,\mu} |\psi^{(1)}_{\eta}\rangle \langle \psi^{(1)}_{\mu}|+\alpha_{g} |g\rangle \langle g|.
\label{dvpt}
\end{align}

The coefficients $\alpha_{\eta,\mu}$ and $\alpha_g$ in the expansion of Eq.\,(\ref{dvpt}) can be computed numerically. Alternatively, these coefficients can also be deduced from the eigenstate decay structure discussed in Sect.\,\ref{sec_dlow}. There, we have seen that -- for the waveguide configuration -- a subradiant two-excitation eigenstate $\ket{\psi_{\xi}^{(2)}}$ (of decay rate $\Gamma_\xi^{(2)}$) decays into its two single-excitation constituents $\xi'_1$ and $\xi'_2$ at rates $\Gamma_{\xi'_2}^{(1)}$ and $\Gamma_{\xi'_1}^{(1)}$, respectively. Thus, only the coefficients for these two single-excitation  states $\alpha_{\xi'_1,\xi'_1}$ and $\alpha_{\xi'_2,\xi'_2}$ must be (significantly) non-zero in the ansatz Eq.\,(\ref{dvpt}). The transition rate to the single-excitation state $\xi'$, following from a Liouvillian eigenstate expansion, is given by $\gamma_{\xi,\xi'}=-\alpha_{\xi',\xi'}\,(\Gamma_\xi^{(2)}-\Gamma_{\xi'}^{(1)})$. That relation combined with the actual transition rates found earlier ($\gamma_{\xi,\xi'_1}=\Gamma_{\xi'_2}^{(1)}$, $\gamma_{\xi,\xi'_2}=\Gamma_{\xi'_1}^{(1)}$) and the property $\Gamma_{\xi}^{(2)}\simeq \Gamma_{\xi'_1}^{(1)}+\Gamma_{\xi'_2}^{(1)}$, suggests that $\alpha_{\xi'_1,\xi'_1}=\alpha_{\xi'_2,\xi'_2} \sim -1$, and similarly $\alpha_g \sim 1$. Indeed by numerically calculating the Liouvillian eigenstates one finds that the coefficients tend to these values with increasing atom number. For the free-space setup, further non-zero coefficients associated with \tblue{superradiant} components exist in addition to the two dominant coefficients $\alpha_{\xi'_1,\xi'_1}$ and $\alpha_{\xi'_2,\xi'_2}$, in line with the existence of decay channels via \tblue{superradiant} states found earlier (e.g., see dotted arrow in Fig.\,\ref{fig:structure}(b)). We compare the transition rates of the most subradiant two-excitation  eigenstate as obtained by both the rate model of Sect.\,\ref{sec_jumppic} and a Liouvillian eigenstate expansion in Fig.\,\ref{Fig3}, which show excellent agreement and further confirm the rate model approximation.

In \ref{Eigenvalues_Liouvillian} we provide additional details about the diagonalization procedure and the {eigenstates} of the Liouvillian $\mathcal{L}$. As {explicitly} illustrated above for one and two excitations, the eigenstate $Z_\Lambda$ of the Liouvillian with eigenvalue $\Lambda=-\Gamma^{(m_{\rm ex})}_{\xi}$ can be constructed by considering the dynamics of the system initially prepared in the density matrix $\rho^{(m_{\rm ex})}_{\xi}=|\psi^{(m_{\rm ex})}_{\xi}\rangle \langle \psi^{(m_{\rm ex})}_{\xi}|$. In addition to $\rho^{(m_{\rm ex})}_{\xi}$, the eigenstate involves terms corresponding to a smaller number of excitations of the form $|\psi^{(n_{\rm ex})}_{\eta}\rangle \langle \psi^{(n_{\rm ex})}_{\mu}|$ with $n_{\rm ex} < m_{\rm ex}$ [see the explicit construction in \ref{Eigenvalues_Liouvillian}], which are populated due to quantum jumps $\mathcal{J}$. In the next subsection, we will refer to such an eigenstate as an $m_{\rm ex}$-excitation eigenstate of the Liouvillian.

\subsection{Power-law behavior from single-excitation Liouvillian eigenstates\label{sec:power_law_from_Liouvillian}}

Any initial density matrix $\rho_0$ at time $t=0$ can be decomposed in terms of the eigenstates $Z_{\Lambda}$ of the Liouvillian. The expectation value of any operator $A$ can thus be written in the form
\begin{align}
\left\langle A \right\rangle(t) =\sum_{\Lambda \in sp(\mathcal{L})} \alpha_{\Lambda}e^{\Lambda t} \textrm{Tr}\left(A Z_{\Lambda}\right).
\label{decomposition_A}
\end{align}
Here, $\Lambda$ are the eigenvalues of the Liouvillian associated with the eigenstates $Z_{\Lambda}$. The coefficients $\alpha_{\Lambda}$ denote the ``overlap'' between $\rho_0$ and the eigenstates $Z_\Lambda$. More specifically, one has $\alpha_{\Lambda}=\textrm{Tr}\left(X_{\Lambda}^{\dagger}\rho_0\right)$, where $X_{\Lambda}$ is the eigenstate of the adjoint of the Liouvillian with eigenvalue $\Lambda^*$ [see \ref{Eigenvalues_Liouvillian}]. At long times, the dynamics is dominated by subradiant eigenstates {(corresponding to small negative real components of $\Lambda$) } as radiant components progressively disappear. 

In the waveguide setup, we find that the long-time dynamics of the population $\langle \hat{n}_e \rangle$ is fully determined by the single-excitation {eigenstates} of the Liouvillian when $N$ is large. This property can be understood by examining the coefficients {$\alpha_{\eta,\mu}$} in the expression of two-excitation subradiant eigenstates $Z$ in Eq.\,(\ref{dvpt}). These coefficients tend to zero when $N$ grows, except two of them which tend to $\alpha_{\xi'_1,\xi'_1}=\alpha_{\xi'_2,\xi'_2} \sim -1$. This leads to $\textrm{Tr}\left( \hat{n}_e Z \right)\simeq 0$, as the contributions of the two-excitation component and the single-excitation components of $Z$ compensate. This property actually extends to all the multi-excitation subradiant eigenstates of the Liouvillian, and one finds that the $m_{\rm ex}$-excitation subradiant eigenstates of the Liouvillian do not contribute to the population dynamics for $m_{\rm ex}>1$. As a result, the long-time dynamics can be simply written 
\begin{align}
&\left\langle \hat{n}_e \right\rangle(t) \sim \sum_{\xi \in~ 1~ \textrm{exc}} \alpha_{\xi} \exp \left[-\Gamma^{(1)}_{\xi} t\right],
\label{decomposition_sigma_z_simplified}
\end{align}
where we wrote for simplicity $\alpha_{\xi}=\alpha_{-\Gamma^{(1)}_{\xi}}$. The dynamical behavior of $\langle \hat{n}_e  \rangle$ at long times arises purely from single-excitation decay rates. It is important to note that this formula fully takes into account the whole many-body dynamics, even if multi-excitation components are not present in an explicit manner. The many-body aspect and its related complexity are encapsulated in the amplitudes $\alpha_{\xi}$. In particular, while only single-excitation eigenstates $Z_{\Lambda}$ contribute to the population $\langle \hat{n}_e \rangle$, the associated eigenstates $X_{\Lambda}$ needed to calculate $\alpha_{\xi}$ contain states up to $N$ excitations. We can compute numerically these amplitudes $\alpha_{\xi}$ at small atom number for an initially fully excited state. We then find a smooth distribution for the most subradiant states, which becomes more and more flat as $N$ increases [see \ref{Eigenvalues_Liouvillian} for details]. Supposing that this distribution becomes uniform at large $N$ and taking the continuum limit in Eq.\,(\ref{decomposition_sigma_z_simplified}), one can estimate the behavior of the population as $\left\langle \hat{n}_e  \right\rangle (t) \sim (\Gamma_{0} t)^{-\eta}$ {when $\Gamma_{0} t \gg 1$}, with $\eta =  0.5$, in accordance with the results of Sec. \ref{sec:power_law_observation}.



\section{Consequences of open quantum criticality on a lattice clock protocol \label{sec:clock}}

We have explained above how the \tbl{algebraic relaxation of the population $\langle \hat{n}_e  \rangle$ of the 1D atomic array can be understood in terms of open critical dynamics.} In this section, we analyze the decay dynamics of the clock signal in a lattice clock protocol, and show similarly that the dynamics is strongly affected by the existence of long-lived subradiant modes. We find in particular that subradiant states induce a time-dependent shift in the measured value of the atomic frequency. At long times, this shift is determined by the most subradiant modes of the system. Furthermore, the size of the clock signal itself exhibits a slow non-exponential decay in time, which allows to extend the clock interrogation time and thus might improve the clock sensitivity beyond standard limits. \tbl{However, the clock signal does not exhibit robust power law behavior, and in fact decays faster than might be expected compared to the excited-state population. \tblue{We provide numerical evidence that the origin of the clock signal decay at long times originates from an effective many-body dephasing, induced by coherent dipole-dipole interactions.}}


\begin{figure}[h!]
\center
\includegraphics[scale=0.42]{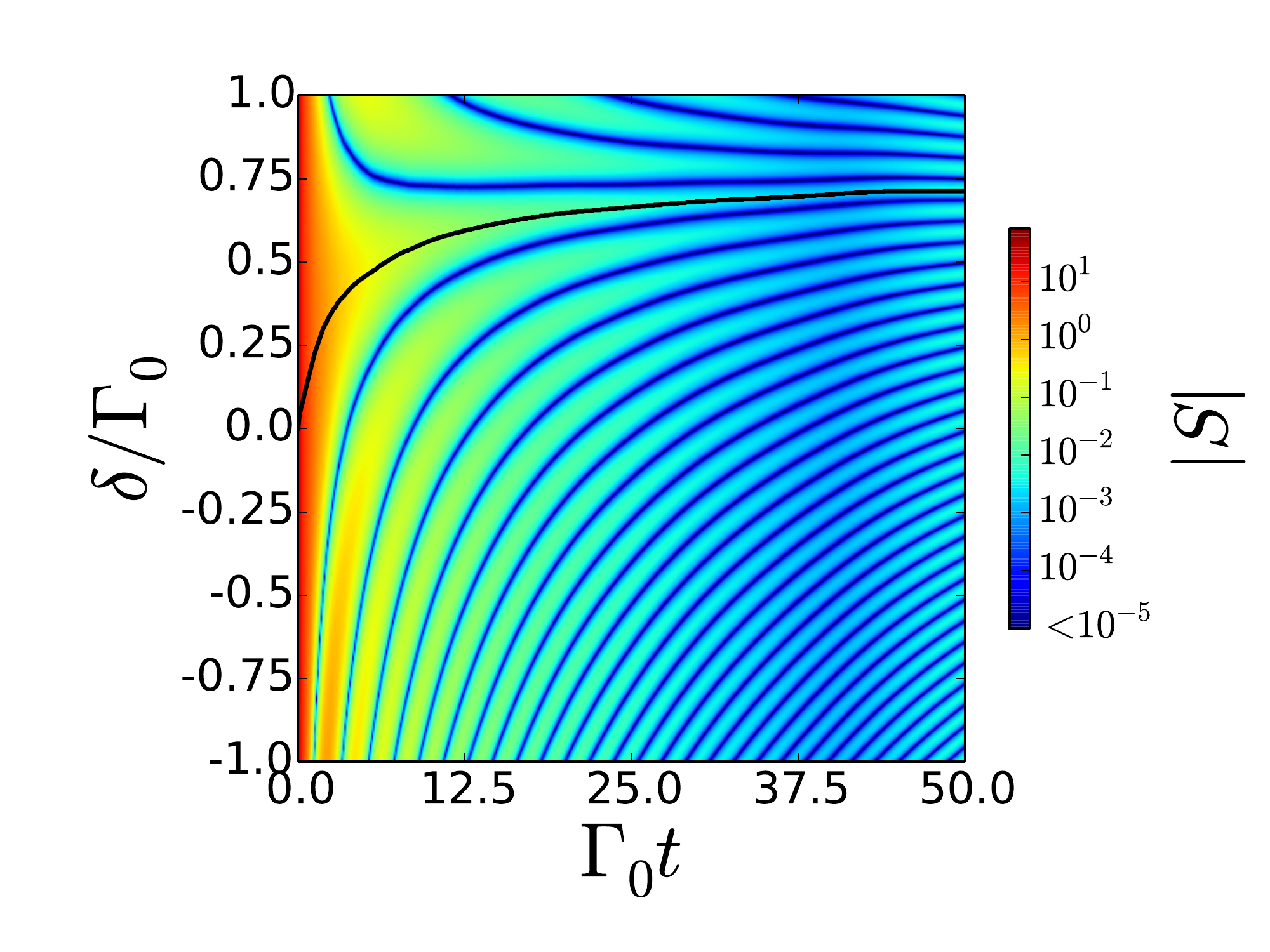}
\caption{\label{Fig_clock} Evolution of clock signal $|S|$ with respect to time and detuning {(in units of $\Gamma_0$)} in a free space configuration, starting from the initial wavefunction  $\bigotimes_{n=1}^N \frac{|g_{n} \rangle+e^{in\pi} |e_{n}\rangle  }{\sqrt{2}}$,  $N=14$ atoms and lattice constant $k_{0}\,d = 0.3\pi$. These results correspond to an evolution of the wave function under stochastic quantum jumps. The {black curve} shows the peak of the central fringe, from which the apparent atomic frequency is obtained. }
\end{figure}

\begin{figure}[h!]
\center
\includegraphics[scale=0.38]{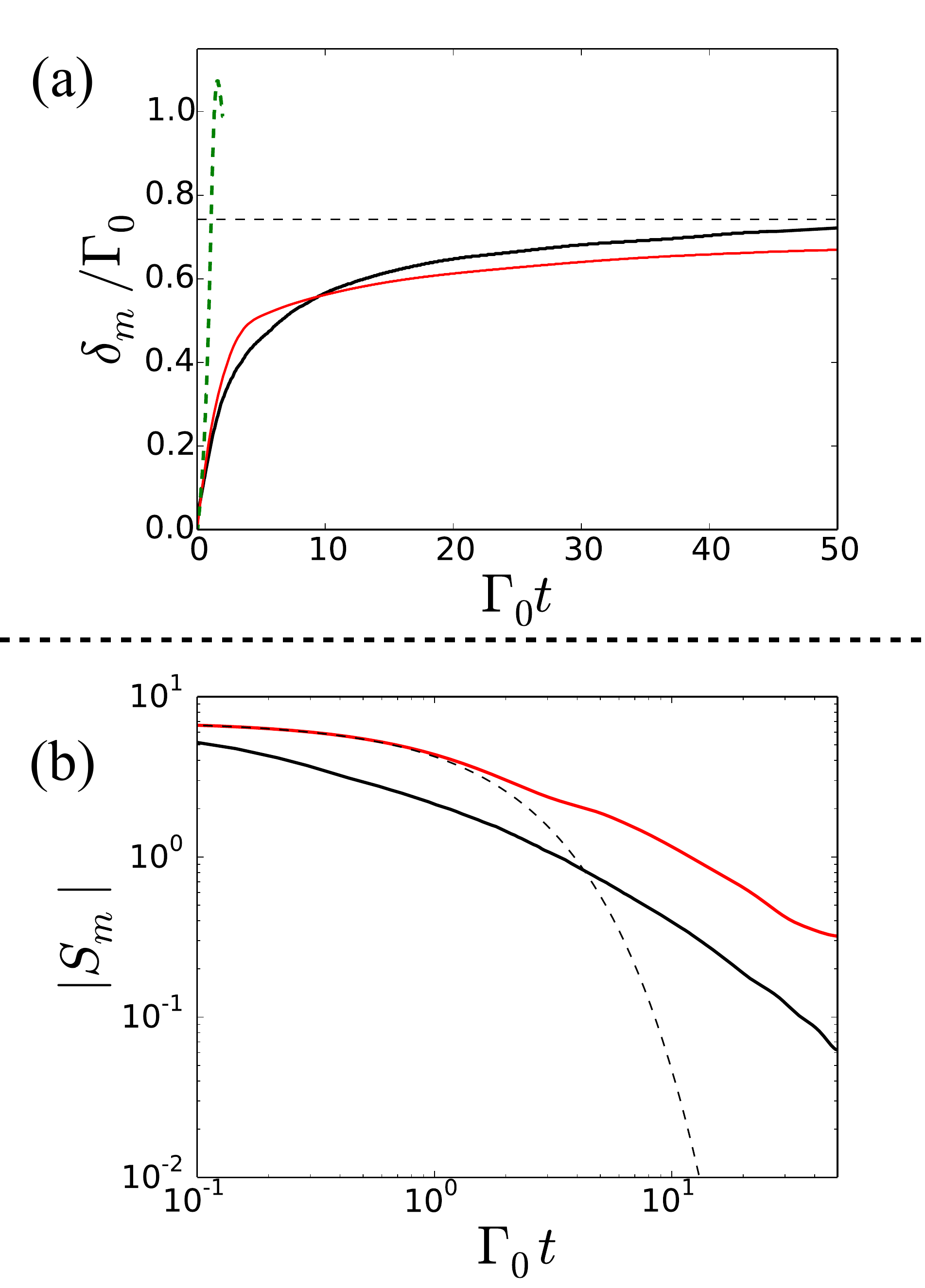}
\caption{\label{Fig_clock_2} (a) Time evolution of the dynamical shift of the central Ramsey fringe $\delta_m$ obtained with three different approaches. The solid back curve corresponds to the evolution of the wave function under stochastic quantum jumps (similarly to Fig.\,\ref{Fig_clock}). The red and dashed green curve correspond, respectively, to a mean field approach taking into account correlations up to second order, and to the short time approach developed in Ref.\,\cite{darrick04}. The horizontal dashed black line denotes the energy shift of the most-subradiant single-excitation eigenstate. We have $N=14$ and $k_{0}\,d = 0.3\pi$. (b) The solid black curve shows the evolution of $|S_m|$ at the center of the central fringe obtained with the stochastic wave function approach. The solid red and dashed black curve correspond, respectively, to the same quantity taking into account correlations up to second order, and in the case of non-interacting atoms.  }
\end{figure}

We analyze the Ramsey spectroscopy protocol previously introduced in Sec.\,\ref{Sect2}, and in particular, the resulting clock signal  $S=\Re\left\langle \sum_{m=1}^N e^{ik_L m d} \sigma_{eg}^m \right\rangle$, where the average is taken just before the second $-\pi/2$ pulse \cite{darrick04}. We show in Fig.\,\ref{Fig_clock} the typical Ramsey fringes obtained for $S$ as a function of time and detuning, for an atomic chain of $N = 14$ atoms in free space, with $k_{0}\,d = 0.3\pi$ and $k_L d=\pi$ {(see Eq.\,(\ref{eq:initial_clock_state}))}. These results were obtained from an evolution of the wave function under stochastic quantum jumps using an average over $10^4$ trajectories. {The black curve} denotes the center of the central fringe $\delta_m$, used to reference the laser frequency. For independent atoms, this would correspond to the line $\delta_m=\omega_L-\omega_0=0$, such that the laser would be {referenced} to the true atomic resonance frequency. We find here that the central fringe is shifted dynamically towards positive detunings $\delta_m>0$, due to the effect of subradiant states. We plot in Fig.\,\ref{Fig_clock_2} (a) the evolution of $\delta_m$ (full black line) and we find that it approaches at long times the frequency shift  $\omega_{1}^{(1)}$ of the most subradiant single-excitation eigenstate $\ket{\psi_{1}^{(1)}}$ denoted by the dashed black curve. (Recall that this state is an eigenstate of $\mathcal{H}_{\rm eff}$ in Eq.\,(\ref{eq:Effective_Hamiltonian}), with complex eigenvalue $\omega_{1}^{(1)}-i\Gamma_{1}^{(1)}/2$.) 

We can also compare our result to the predictions obtained by previous perturbative theoretical approaches, such as the short-time expansion of Ref.\,\cite{darrick04}, or mean field like methods\,\cite{kramer15}. In Fig.\,\ref{Fig_clock_2}\,(a), the predicted shift of the short-time expansion is shown in green, while the shift {obtained by} a second-order mean field theory (see Sec.\,\ref{sec:power_law_observation} for details) is shown in red. Both approaches quantitatively agree at short times, before correlations build up in the system. The shift predicted by second-order mean field theory qualitatively agrees with the full result, although it predicts a significantly larger signal amplitude. This is illustrated in Fig.\,\ref{Fig_clock_2}\,(b) where we show the maximal signal amplitude $|S_m|$ {along the central fringe as a function of time}, both with the second-order mean field approach (solid red curve) and the full result obtained with the exact stochastic wavefunction approach (solid black curve{, i.e. following the solid black curve in Fig.\,\ref{Fig_clock}}). \tblue{The larger value of the signal amplitude obtained at long times} with the approximate mean-field approach can be understood physically by inspecting the state of the system after the first $\pi/2$ pulse [see Eq.\,(\ref{eq:initial_clock_state}) with $k_L d=\pi$]. This state already contains {the} phase relationship between excited atoms corresponding to subradiance, whose {effect} is enhanced by the mean-field approximation at long times. In contrast, in the \tbl{exact solution}, the population in the subradiant states \textit{dynamically} builds up, following decay from highly excited states. 

In Fig.\,\ref{Fig_clock_2} (b), we also show the evolution of the maximal signal amplitude $|S_m|$ at the center of the brightest fringe in the case of independent atoms (dashed black line). At short times, the maximal signal of the full result decays faster due to the presence of superradiant states. In contrast, at longer times, subradiant states are predominant and one observes a clear non-exponential decay of the signal amplitude. 

\begin{figure}[h!]
\center
\includegraphics[scale=0.38]{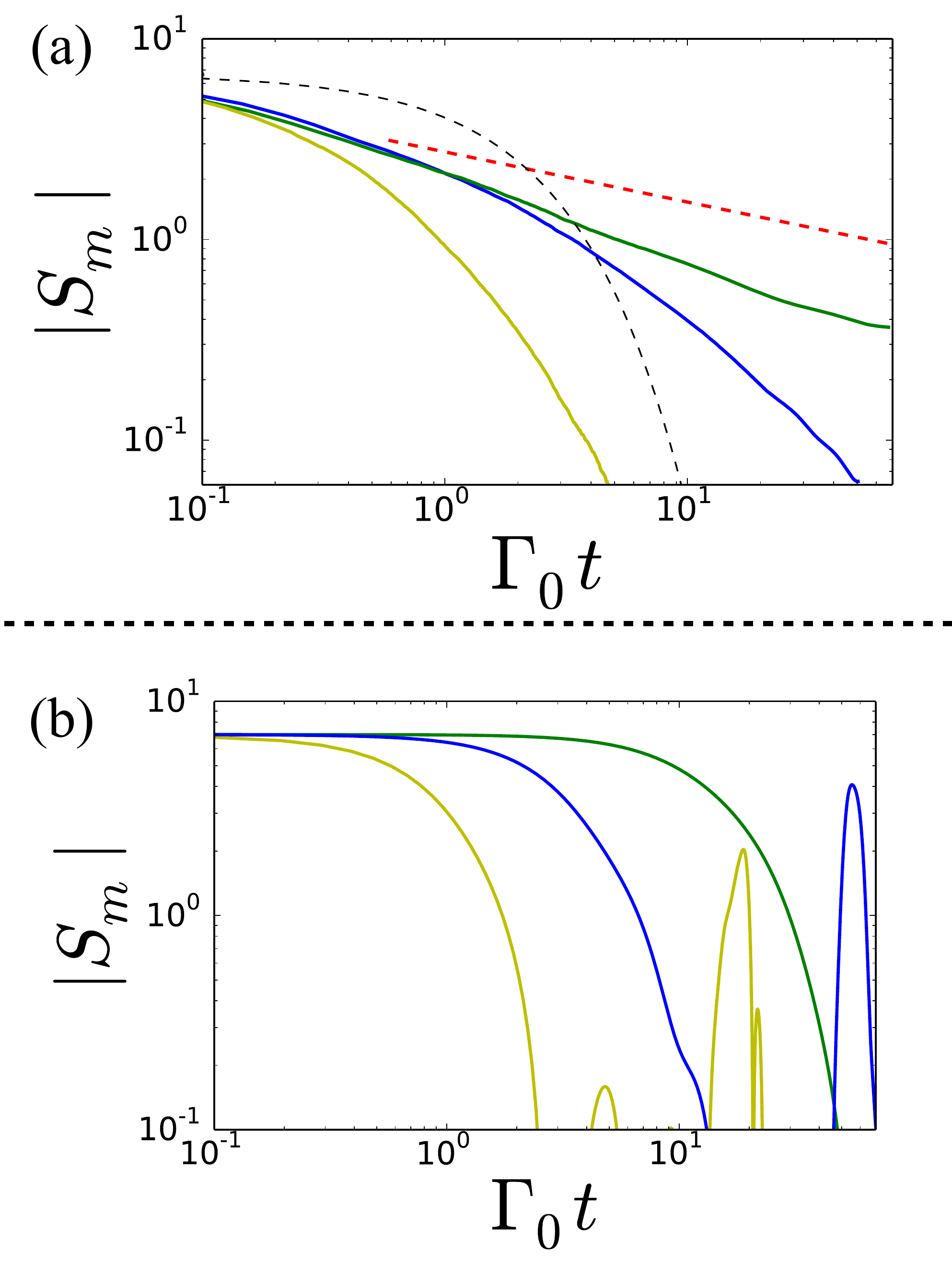}
\caption{\label{Fig_clock_3} (a) Amplitude of the signal at the center of the brightest fringe for a chain of $N=14$ atoms in free space, obtained from an evolution of the wave function under stochastic quantum jumps, for $d/\lambda_0=0.2$ (yellow), $d/\lambda_0=0.3$ (blue), $d/\lambda_0=0.4$ (green). The dashed black curve corresponds to the case of non-interacting atoms. The dashed red line shows a power-law guide to the eye with exponent $\nu=0.25$.  (b) Amplitude of the signal at the center of the brightest fringe for a chain of $N=14$ atoms in free space, obtained from an evolution under the coherent part of the effective Hamiltonian $\mathcal{H}_{\rm eff}$ only, {for $d/\lambda_0=0.2$ (yellow), $d/\lambda_0=0.3$ (blue) and $d/\lambda_0=0.4$ (green)}. A faster initial decay of the signal amplitude is observed for smaller inter-atomic distances $d$. We also note the existence of signal revivals at long times (see {yellow and blue curves}) due to absence of loss. }
\end{figure}

In Fig.\,\ref{Fig_clock_3}(a), we plot the long-time dynamics of the signal amplitude at the center of the brightest fringe, for selected values of $d/\lambda_0$. No robust power law decay behavior is observed, in contrast with the excited population. Furthermore, given a power law with coefficient $\eta=0.5$ for the population, the most naive expectation would be that the clock signal (involving atomic coherence rather than population) might exhibit a power law decay of $\nu=\eta/2$. However, it is seen that the instantaneous slope of the clock signal (on this log-log scale) generally exceeds $\nu> 0.25$ (in absolute value), indicating a faster-than-expected decay. 

To partially understand the lack of a power law, first, we confirm that there exists no analogous picture of the clock signal dynamics in terms of single-excitation Liouvillian eigenstates, as was developed for the population in Sec.\,\ref{sec:power_law_from_Liouvillian}. In particular, the clock signal at long times contains contributions from long-lived higher excitation eigenstates, and without a clear distribution pattern (at least at the system sizes we consider) from which Eq.\,(\ref{decomposition_A}) might be approximately evaluated. \tblue{At an intuitive level, we hypothesize that as the clock signal depends on the sum of single-atom coherences $\sigma_{ge}$, it is thus susceptible to an effective many-body dephasing, which arises from the coherent (Hermitian) part of the dipole-dipole interaction Hamiltonian, Eq.\,(\ref{eq:Effective_Hamiltonian}). These interactions scramble the identical precession that the individual atomic dynamics would undergo on a Bloch sphere (see Fig.\,\ref{FigRamsey}), in the non-interacting case.}

\tblue{In order to check this hypothesis, in Fig.\,\ref{Fig_clock_3}(b) we plot the clock signal amplitude for the same lattice constants as in Fig.\,\ref{Fig_clock_3}(a), but now considering time evolution under the Hermitian part of $\mathcal{H}_{\rm eff}$ only, with no atomic decay processes. One sees that the coherent interactions themselves result in a signal decay (with revivals at long times due to finite size and absence of losses). Furthermore, both in Figs.\,\ref{Fig_clock_3}(a) and (b), the decay becomes slower with increasing lattice constant $d$, even though this results in fewer subradiant states, which we attribute to the strongly decreasing magnitude ($\sim 1/d^3$) of the coherent interactions for atomic transition dipole moments aligned along the axis of the chain. As a final check, we can also repeat these calculations for the case of atomic dipole moments oriented perpendicular to the chain axis (see\,\ref{appendix:perpendicular}). This configuration results in relatively flat band structure for single-excitation subradiant states, as compared to the case of parallel orientation, such that dephasing arising from differential energy shifts should be expected to play a smaller role. Indeed, in this case, we find that a decay more closely resembling a power law emerges, and with an instantaneous slope of $\nu \approx 0.25$ (in absolute value) that more directly reflects the decay of the excited state population itself.}

\tbl{In general, the ability to extend the interrogation time without experiencing exponential losses should be beneficial for clock sensitivity. }
\tbl{A particularly interesting limit is in the window of time evolution where the instantaneous slope of the decay has a value $\nu<0.5$. In that case, the decay in signal over a single interrogation is slower than the typical $\Delta\omega\propto \sqrt{1/T_{\rm avg}}$ scaling obtained by making many independent interrogations over a total averaging time $T_{\rm avg}$. In that case, the optimal clock protocol, absent any other imperfections, would be to run a single interrogation over the entire time $T_{\rm avg}$.} 



\section{Atoms in a 3D lattice \label{sec:3D}}

\tblue{In the case of a 1D array, the emergence of highly subradiant states and the closing of the Liouvillian gap only occur for lattice constants $d<\lambda_0/2$, which are not readily generated by conventional optical lattices. However, here we show that in a 3D lattice, the closing of the Liouvillian gap occurs even for lattice constants $d>\lambda_0/2$. While in 3D, full simulations of the master equation (\ref{eq:Liouville_master_equation}) are limited to too small system sizes\,\cite{Maier14} to extrapolate any behavior in the thermodynamic limit, the closing gap at least strongly suggests that realistic clocks might exhibit similar critical slow-down dynamics as found in 1D.}

\begin{figure}[h!]
\center
\includegraphics[scale=0.45]{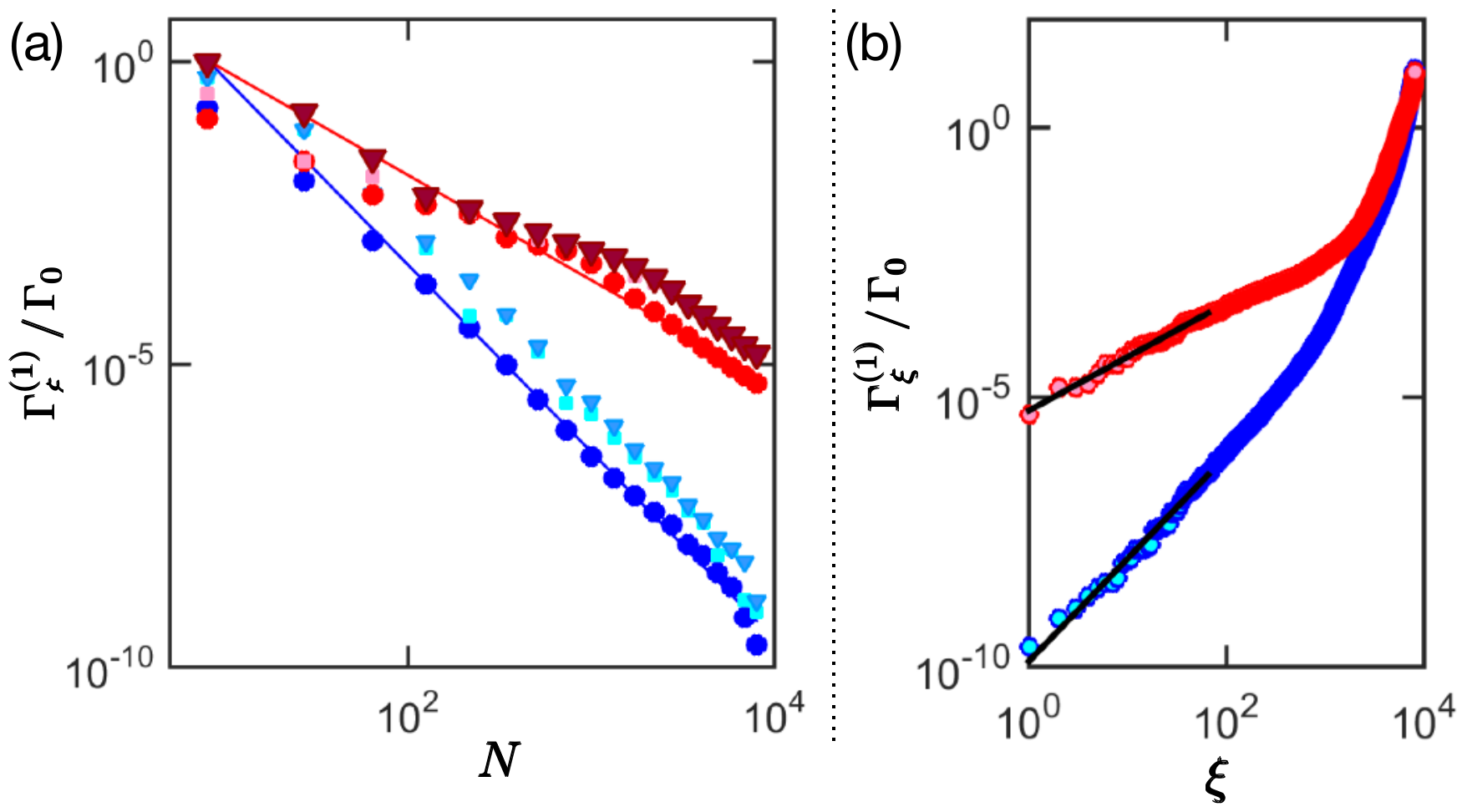}
\caption{\label{Fig3D} Single-excitation decay rate of eigenstate $\xi$ (numbered by increasing decay rate) for a 3D-cube of atoms with lattice constant $d=0.4\lambda_0$ (blue) and $d=0.6\lambda_0$ (red). (a) Decay rate scaling with atom number $N$ for the three most subradiant eigenstates $\xi=1$ (circles), $\xi=2$ (squares) and $\xi=3$ (triangles). Lines represent linear fits to a scaling $\Gamma_\xi^{(1)}\sim N^{-\alpha}$, where $\alpha \simeq 3.1$ and $\alpha\simeq 1.7$ for $d=0.4\lambda_0$ and $d=0.6\lambda_0$, respectively. (b) Single atom decay rate scaling with the numbering index $\xi$ for a 3D cube of $20\times20\times 20$ atoms. Black lines are linear fits to the scaling $\Gamma_\xi^{(1)}\sim \xi^\beta $ for small $\xi$ ($\xi\leq 60$) with $\beta\simeq 1.9$ and $\beta\simeq 1.0$ for $d=0.4\lambda_0$ and $d=0.6\lambda_0$, respectively. }
\end{figure}

To demonstrate a smooth spectrum of decay rates and a closing Liouvillian gap, it is sufficient to consider the single-excitation manifold. In particular, we consider a 3D cube of $N$ two-level atoms (with $N^{1/3}$ sites in each direction), with the axes of the cube aligned along $\hat{x},\hat{y},\hat{z}$ and the atomic dipole moment $p$ along $\hat{z}$. The Green's function between any two lattice sites, projected along the dipole direction, is given by ${\bf{p}}^\dagger \doublearrow{{\bf{G}}}({\bf{r}}_n,{\bf{r}}_m,\omega_0){\bf{p}}=|\mathbf{p}|^2 e^{i k_0 r}/(4 \pi k_0^2 r^3)\,(k_0^2 r^2+i k_0 r-1+z^2 [-k^2-i3k/r+3/r^2])$, where $r=|\mathbf{r}_n-\mathbf{r}_m|$ and $z=z_n-z_m$. We then diagonalize the effective Hamiltonian $\mathcal{H}_{\rm eff}$ of Eq.\,(\ref{eq:Effective_Hamiltonian}) within the single-excitation manifold, and obtain the decay rate spectra $\Gamma_{\xi}^{(1)}$. In Fig.\,\ref{Fig3D}\,(a), we plot the scaling of $\Gamma_{\xi}^{(1)}$ with $N$ for the few most subradiant eigenstates ($\xi=1,2,3$), and for two different lattice constants $d=0.4\lambda_0$ and $d=0.6\lambda_0$. These decay rates are seen to decrease polynomially as $N^{-\alpha}$, where $\alpha$ varies depending on the lattice constant. Moreover, for a fixed atom number, decay rates of eigenstates are smoothly distributed -- shown in Fig.\,\ref{Fig3D}\,(b), where the decay rate is plotted as a function of the eigenstate numbering coefficient $\xi$. The scaling with $\xi$, $\Gamma_{\xi}^{(1)}\propto \xi^{\beta}$, is seen to depend as well on the lattice constant $d$, unlike the 1D case. However, the analysis here is restricted to a rather small maximum cube size of $20\times 20\times 20$ atoms, such that the effect of boundaries might be crucial and a potential universal behavior not reached yet. 


\section{Conclusion and outlook}

We have shown that an optical lattice clock composed of atoms in a 1D array \tbl{exhibits critical open system dynamics}, due to the presence of a continuum of subradiant many-excitation states whose decay rates approach zero in the large array limit. This system exhibits a number of interesting characteristics, including a power-law decay of observables at long times, the growth of ``fermionic'' spatial correations between excitations, and a time-dependent shift of the clock frequency that goes toward the interaction energy of subradiant states. These features cannot be captured by mean-field theories, indicating that such a system is strongly correlated. 

While our analysis was restricted to 1D, we have also \tbl{shown} that actual 3D lattice clocks exhibit \tbl{one of the} key ingredients for open critical behavior, in particular, the continuum of decay rates approaching zero. From a theoretical standpoint, the 3D case seems to present a considerable analytical and numerical challenge to solve, and we anticipate that our results could spark interest in this problem, particularly given the growing general interest in quantum many-body open systems\,\cite{Bohnet16,luschen17,sieberer16,fossfeig17}. Our results could stimulate new experimental directions as well. In regard to actual clock platforms, our work could stimulate novel many-body directions to investigate. While the direct observation of subradiant dynamics could be hard given the lifetimes of some of the proposed transitions (e.g., $\Gamma_0^{-1}\approx 150\,{\rm s}$ on the $^1S_0-^3P_0$ transition of $^{87}$Sr \,\cite{ludlow15, Katori03}), they could be immediately feasible using somewhat faster transitions (e.g., $^1S_0-^3P_1$ in $^{87}{\rm Sr}$, which has a lifetime of $\sim 20\mu{\rm s}$\,\cite{ludlow15}). The possibility of greatly extending excited-state lifetimes through subradiance might also enable a much larger class of atoms and transitions to be used for clocks, whose individual lifetimes are nominally too short to make clock technology feasible. Moreover, given that many-body effects can already be seen for small numbers ($N\sim 10$) of atoms, it might be interesting to explore such dynamics in emerging systems of optical tweezer arrays\,\cite{barredo16,Endres16}. This could enable phenomena such as fermionic correlations to be investigated in atom-by-atom detail.

Finally, beyond specific application to clocks, our work provides the first comprehensive picture of subradiance in arrays of atoms at the many-body level, complementing the growing body of work that already demonstrates rich behavior at the level of single excitations\,\cite{Facchinetti16,Plankensteiner15,Bettles15,Bettles16a,Bettles16b,plankensteiner17}. For example, at the level of single excitations, it has already been shown that subradiance can enable reduced errors for applications such as quantum memories\,\cite{asenjo17,manzoni18} and allow for topological edge states in two-dimensional arrays\,\cite{perczel17, bettles17}. Our work provides critical insight to extend the use of subradiance generally to applications involving many excitations, and to investigate the effect of interactions between topological excitations.

\begin{acknowledgments}
D.E.C. acknowledges support from Fundacio Privada Cellex, Spanish MINECO Severo Ochoa Program SEV-2015-0522, MINECO Plan Nacional Grant CANS, CERCA Programme/Generalitat de Catalunya, AGAUR Grant 2017 SGR 1334 and ERC Starting Grant FOQAL.
\end{acknowledgments}

\section*{Appendices}

\setcounter{figure}{0} 
\setcounter{section}{0}
\setcounter{equation}{0}
\renewcommand{\thefigure}{A\arabic{figure}}
\renewcommand{\thesection}{Appendix \Alph{section}}
\renewcommand{\thesubsection}{\arabic{subsection}}
\renewcommand{\theequation}{A\arabic{equation}}

\section{ Eigenelements of the Liouvillian }
\label{Eigenvalues_Liouvillian}
In this Appendix, we describe how to construct eigenstates of the Liouvillian of Eq.\,(\ref{eq:Liouville_master_equation}), which gives the density matrix evolution under dipole-dipole interactions. We first write the Liouvillians $\mathcal{L}= \mathcal{K}+\mathcal{J}$, where 

\begin{align}
& \mathcal{K}[\rho]=\frac{1}{i\hbar}(\mathcal{H}_{\rm eff} \rho-\rho \mathcal{H}_{\rm eff}^{\dagger}),\label{Liouville_eq_part_1}\\
 &\mathcal{J}[\rho]=\sum_{m,n}\Gamma_{m,n}\sigma^m_{ge}\rho \sigma^n_{eg}.\label{Liouville_eq_part_2}
\end{align}
The effective Hamiltonian $\mathcal{H}_{\rm eff}$ commutes with $\hat{n}_e =\sum_{n}\sigma^n_{ee}$, so that one can look for its eigenstates within a given excitation manifold. As $\mathcal{H}_{\rm eff}$ is complex symmetric, it can be diagonalized in a complex orthogonal basis \cite{Horn_johnson}. We denote by $|\psi_{\xi}^{(m_{\rm ex})} \rangle$ the right eigenstates of $\mathcal{H}_{\rm eff}$ with $m_{\rm ex}$ excitations, and $\lambda_{\xi}^{(m_{\rm ex})}=\hbar(\omega^{(m_{\rm ex})}_{\xi}-i\Gamma^{(m_{\rm ex})}_{\xi}/2)$ the corresponding eigenvalue. Here $\omega^{(m_{\rm ex})}_{\xi}$ and $\Gamma^{(m_{\rm ex})}_{\xi}>0$ physically represent the renormalized frequency and decay rates associated with $|\psi^{(m_{\rm ex})}_{\xi}\rangle$. Here, the index $\xi$ runs from 1 to $d_{m_{\rm ex}}=\binom{N}{m_{\rm ex}}$.\\

 We next expose the different steps to diagonalize the Liouvillian $\mathcal{L}$, following Refs.\,\cite{briegel_englert,Barnett_stenholm,Torres}. We first explicitly build the eigenstates of $\mathcal{K}$ from the right eigenstates of $\mathcal{H}_{\rm eff}$. We define
\begin{align}
U^{(n_{\rm ex},l_{\rm ex})}_{\xi_1,\xi_2} =|\psi_{\xi_1}^{(n_{\rm ex}+l_{\rm ex})} \rangle \langle \psi^{(n_{\rm ex})}_{\xi_2}|,\label{def_U}
\end{align}
where $n_{\rm ex},n_{\rm ex}+l_{\rm ex} \in \{0,1,..,N\}$ number the excitation manifold of the corresponding vector. We can check that $U^{(n_{\rm ex},l_{\rm ex})}_{\xi_1,\xi_2}$ are eigenstates of $\mathcal{K}$, with
\begin{align}
\mathcal{K}U^{(n_{\rm ex},l_{\rm ex})}_{\xi_1,\xi_2} &=\frac{\lambda_{\xi_1}^{(n_{\rm ex}+l_{\rm ex})}-\left[\lambda_{\xi_2}^{(n_{\rm ex})}\right]^*}{i\hbar}U^{(n_{\rm ex},l_{\rm ex})}_{\xi_1,\xi_2} \label{K_on_U}\notag \\
&\equiv\Lambda^{(n_{\rm ex},l_{\rm ex})}_{\xi_1,\xi_2} U^{(n_{\rm ex},l_{\rm ex})}_{\xi_1,\xi_2}.
\end{align}
The eigenvalues $\Lambda^{(n_{\rm ex},l_{\rm ex})}_{\xi_1,\xi_2}$ have a negative real part $-\left(\Gamma_{\xi_1}^{(n_{\rm ex}+l_{\rm ex})}+\Gamma_{\xi_2}^{(n_{\rm ex})}\right)/2$ corresponding to the sum of the decay rates associated with states $|\psi_{\xi_1}^{(n_{\rm ex}+l_{\rm ex})} \rangle$ and $|\psi_{\xi_2}^{(n_{\rm ex})} \rangle$. \tblue{This real part tends to zero in the thermodynamic limit $N\to \infty$ when $\xi_1$ and $\xi_2$ are indices corresponding to strongly subradiant eigenstates, i.e. the Liouvillian gap closes.} 

We next show that the eigenvalues of $\mathcal{L}$ are those of $\mathcal{K}$. To prove this statement, it should be first noted that $\mathcal{J}$ physically lowers the number of excitations of a pure state in the Hilbert space by one. This implies that the operator $\mathcal{J}$ sends a given vector $U^{(n_{\rm ex},l_{\rm ex})}_{\xi_1,\xi_2}$ onto a linear combination (in terms of $\eta_1$ and $\eta_2$) of the vectors $U^{(n_{\rm ex}-1,l_{\rm ex})}_{\eta_1,\eta_2}$ for $n_{\rm ex},n_{\rm ex}+l_{\rm ex}>0$. Note that $\mathcal{J}$ conserves the number $l_{\rm ex}$ and changes $n_{\rm ex}$ to $n_{\rm ex}-1$. 
Consider then that we order the set of vectors $U^{(n_{\rm ex},l_{\rm ex})}_{\xi_1,\xi_2}$ by increasing values of $n_{\rm ex}$. In this basis, $\mathcal{J}$ has a strict triangular form. From that, we conclude that the eigenvalues of $\mathcal{L}$ are the ones of $\mathcal{K}$. We furthermore conclude that if all the eigenvalues of $\mathcal{K}$ are distinct, then $\mathcal{L}$ can be diagonalized. The conservation of $l_{\rm ex}$ (under the action of both $\mathcal{J}$ and $\mathcal{K}$) is related to the fact that the operator $\mathcal{F}[\rho]=\hat{n}_e \rho-\rho \hat{n}_e =[\hat{n}_e ,\rho]$ where $\hat{n}_e =\sum_m \sigma^m_{ee} $, commutes with the Liouvillian, as also noted in Ref.\,\cite{Ribeiro18}.

Having deduced the eigenvalues of $\mathcal{L}$, we can now construct their corresponding eigenstates, based on Refs.\,\cite{briegel_englert,Barnett_stenholm,Torres}. We define $Z_{\xi_1,\xi_2}^{(m_{\rm ex},l_{\rm ex})}$ as the Liouvillian eigenstate associated with the eigenvalue $\Lambda_{\xi_1,\xi_2}^{(m_{\rm ex},l_{\rm ex})}$, i.e.
\begin{equation}\label{LO_eval} \mathcal{L}\,Z_{\xi_1,\xi_2}^{(m_{\rm ex},l_{\rm ex})} = \Lambda_{\xi_1,\xi_2}^{(m_{\rm ex},l_{\rm ex})}\,Z_{\xi_1,\xi_2}^{(m_{\rm ex},l_{\rm ex})}. \end{equation}
Following the previous discussion, such eigenstates can be constructed in subspaces of constant $l_{\rm ex}$. More precisely, one can build eigenstates $Z^{(m_{\rm ex},l_{\rm ex})}_{\xi_1,\xi_2}$ from the states $U^{(n_{\rm ex},l_{\rm ex})}_{\xi_1',\xi_2'}$ of Eq.\,(\ref{def_U}), suggesting an ansatz $Z^{(m_{\rm ex},l_{\rm ex})}_{\xi_1,\xi_2}=\sum_{n_{\rm ex}=1}^N \sum_{\xi_1',\xi_2'} c_{\xi_1',\xi_2'}^{(n_{\rm ex},l_{\rm ex})} U^{(n_{\rm ex},l_{\rm ex})}_{\xi_1',\xi_2'}$. Inserting that ansatz into the eigenvalue equation Eq.\,(\ref{LO_eval}), and using Eq.\,(\ref{K_on_U}), leads to
\begin{align}
 \left[\Lambda^{(m_{\rm ex},l_{\rm ex})}_{\xi_1,\xi_2}-\Lambda^{(n_{\rm ex},l_{\rm ex})}_{\xi_1',\xi_2'}\right]&c_{\xi_1',\xi_2'}^{(n_{\rm ex},l_{\rm ex})}\notag \\
 &=\sum_{\xi_1'',\xi_2''}  J_{(\xi_1'',\xi_2'')\to(\xi_1',\xi_2')}^{(n_{\rm ex}+1,l_{\rm ex})}c_{\xi_1'',\xi_2''}^{(n_{\rm ex}+1,l_{\rm ex})}\,.
\label{construction_eigenstate}
\end{align}
Here, the quantity $J$ describes the action of the jump term $\mathcal{J}$ on $U^{(n_{\rm ex},l_{\rm ex})}_{\xi_1,\xi_2}$,
\begin{align}
\mathcal{J}U^{(n_{\rm ex},l_{\rm ex})}_{\xi_1,\xi_2}=\sum_{\xi_1',\xi_2'} J_{(\xi_1,\xi_2)\to(\xi_1',\xi_2')}^{(n_{\rm ex},l_{\rm ex})} U^{(n_{\rm ex}-1,l_{\rm ex})}_{\xi_1',\xi_2'}.
\label{J_on_U}
\end{align}
The recurrence relation (\ref{construction_eigenstate}) is valid for all $n_{\rm ex}\in \left\{0,1,..,N-l-1\right\}$ and $(\xi_1,\xi_2)$. For $n_{\rm ex}+l_{\rm ex}=N$, the right hand side of Eq.\,(\ref{construction_eigenstate}) is zero. Using successively the relation (\ref{construction_eigenstate}) for decreasing values of $n_{\rm ex}$, we find that $c_{\xi_1,\xi_2}^{(n_{\rm ex},l_{\rm ex})}=0$ if $n_{\rm ex}>m_{\rm ex}$. For $n_{\rm ex}=m_{\rm ex}$, the right hand side of Eq.\,(\ref{construction_eigenstate}) still vanishes but one can choose a non-zero value for $c_{\xi_1,\xi_2}^{(m_{\rm ex},l_{\rm ex})}$ as the difference of eigenvalues $(\Lambda^{(m_{\rm ex},l_{\rm ex})}_{\xi_1,\xi_2}-\Lambda^{(n_{\rm ex},l_{\rm ex})}_{\xi_1',\xi_2'})$ vanishes for $m_{\rm ex}=n_{\rm ex}$ and $(\xi_1,\xi_2)=(\xi_1',\xi_2')$. From there, one finds the other components of the eigenstate. This recursion is not well defined if there exist for the same $l_{\rm ex}$ two different triplets $(m_{\rm ex},\xi_1,\xi_2)\neq(n_{\rm ex},\xi_1',\xi_2')$ such that $\Lambda^{(m_{\rm ex},l_{\rm ex})}_{\xi_1,\xi_2}=\Lambda^{(n_{\rm ex},l_{\rm ex})}_{\xi_1',\xi_2'}$. We recover our criterion for the diagonalization : if the eigenvalues of $\mathcal{K}$ are distinct, then the recursion is well defined. Note that the construction presented here is exactly what has been done in Sect.\,\ref{eigenstate_construction}, starting from 
$U^{(n_{\rm ex},l_{\rm ex}=0)}_{\xi,\xi}$ for $n_{\rm ex}=1,2$.

As illustrated in Ref.\,\cite{briegel_englert}, the knowledge of the right eigenelements of $\mathcal{L}$ is not sufficient to determine the expansion of any density matrix in the basis of right eigenstates. One also needs the eigenstates of the adjoint operator of $\mathcal{L}$ with respect to the usual Hilbert Schmidt (HS) inner product on $\mathcal{B}(\mathcal{H})$, the space of the linear operators on the Hilbert space $\mathcal{H}$. Let us first look for the eigenstates of $\mathcal{K}^{\dagger}$, the adjoint of $\mathcal{K}$. We recall that $\mathcal{K}^{\dagger}$ is the adjoint of $\mathcal{K}$ if we have $\langle A |\mathcal{K} B \rangle_{HS}=\langle \mathcal{K}^{\dagger} A | B \rangle_{HS}$. One finds that $\mathcal{K}^{\dagger}$ is defined by
\begin{align}
\mathcal{K}^{\dagger}\rho=\frac{1}{i\hbar}(\rho\mathcal{H}_{\rm eff} -\mathcal{H}_{\rm eff}^{\dagger}\rho)
\label{adjoint}
\end{align}
 We find the eigenstates of $\mathcal{K}^{\dagger}$ to be of the form $V^{(n_{\rm ex},l_{\rm ex})}_{\xi_1,\xi_2} =|\varphi_{\xi_1}^{(n_{\rm ex}+l_{\rm ex})} \rangle \langle \varphi^{(n_{\rm ex})}_{\xi_2}|$ with $\mathcal{K}^{\dagger}V^{(n_{\rm ex},l_{\rm ex})}_{\xi_1,\xi_2} =\left[\Lambda^{(n_{\rm ex},l_{\rm ex})}_{\xi_1,\xi_2}\right]^*  V^{(n_{\rm ex},l_{\rm ex})}_{\xi_1,\xi_2}$. Here, $\langle \varphi^{(n_{\rm ex})}_{\xi}|$ is the left eigenvector of the effective Hamiltonian associated with $|\psi_{\xi}^{(n_{\rm ex})} \rangle$. Note that we have the following properties, $\langle \varphi^{(n_1)}_{\xi_1}|\psi_{\xi_2}^{(n_2)} \rangle=\delta_{\xi_1,\xi_2}\delta_{n_1,n_2}$ and $\mathds{1}=\sum_{n_{\rm ex}}\sum_{\xi} \ket{\psi_{\xi}^{(n_{\rm ex})}}\bra{\varphi_{\xi}^{(n_{\rm ex})}}$. The diagonalization of the adjoint of the Liouvillian follows from the one of $\mathcal{K}^{\dagger}$, applying the same procedure as the one outlined above for $\mathcal{L}$. In the case of distinct eigenvalues for $\mathcal{K}$, we finally find a complete set of eigenstates $X^{(n_{\rm ex},l_{\rm ex})}_{\xi_1,\xi_2}$ of $\mathcal{L}^{\dagger}$, each associated with one eigenstate $Z^{(n_{\rm ex},l_{\rm ex})}_{\xi_1,\xi_2}$ of $\mathcal{L}$, but with conjugated eigenvalues.  

\begin{figure}[h!]
\includegraphics[scale=0.32]{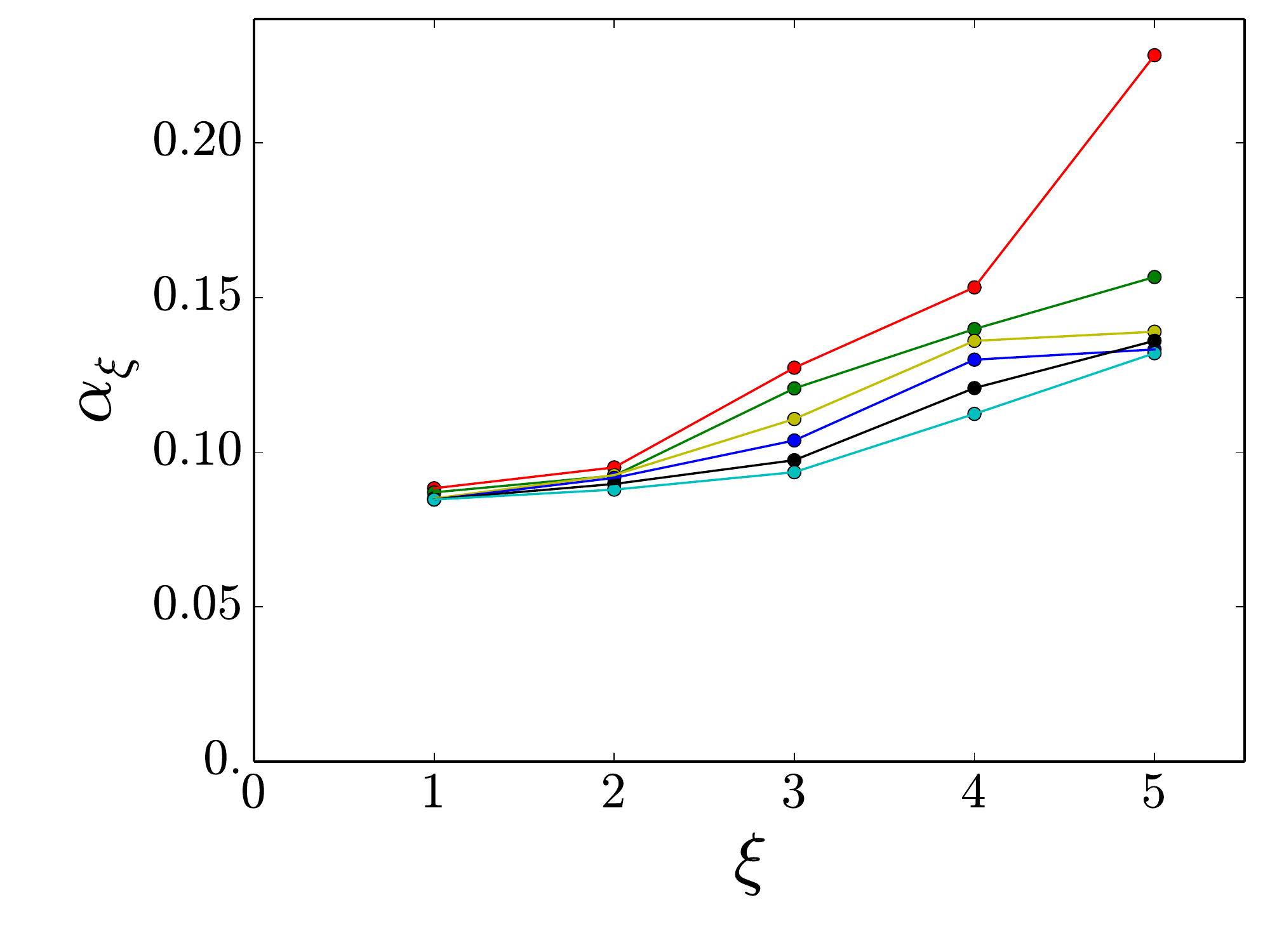}
\caption{\label{lfigcoeffalpha}  Coefficients $\alpha_{\xi}$ for the first five one-excitation sub-radiant eigenstates for $N=8$ (red), $N=9$ (green), $N=10$ (yellow), $N=11$ (blue), $N=12$ (black), $N=13$ (cyan) for $d/\lambda_0=0.1$. }
\end{figure}

The diagonalization of the Liouvillian described above allows us to expand the time-evolved density matrix of the system in the eigenbasis. Starting from any initial density matrix $\rho_0$ at time $t=0$, we have more specifically the unique decomposition,
\begin{align}
&\mathcal{\rho}(t)=e^{\mathcal{L}t}\rho_0=\sum_{\Lambda \in sp(\mathcal{L})} \textrm{Tr}(X_{\Lambda}^{\dagger}\rho_0)e^{\Lambda t}Z_{\Lambda}.
\label{decomposition}
\end{align}
As shown in the main text, the long time behavior of the total population can be written in the waveguide setup as a function of the single-excitation eigenstates only,
\begin{align}
&\left\langle \hat{n}_e  \right\rangle(t) \sim \sum_{\xi} \alpha_{\xi} \exp \left[-\Gamma^{(1)}_{\xi} t\right].
\label{decomposition_p_appendix}
\end{align}
One can compute numerically the coefficients $\alpha_{\xi}=\textrm{Tr}(X_{-\Gamma^{(1)}_{\xi}}^{\dagger}\rho_0)$ after having determined the eigenstates $X_{-\Gamma^{(1)}_{\xi}}$. We show in Fig.\,\ref{lfigcoeffalpha} the values obtained for a fully excited initial state and different atom number. One finds that the distribution of the coefficients $\alpha_{\xi}$ becomes more and more flat as the atom number increases. This flat distribution then allows to estimate the power-law exponent for the population decay at long times.

We have seen that only the single-excitation eigenstates of the Liouvillian contribute to the observable $\hat{n}_e$ at long times. Similarly, it can be shown that the particular decay structure in the waveguide setup implies that only the $r$-excitation eigenstates contribute to the $r$-body observable {$\sum_{n_1,..,n_r} \sigma^{n_1}_{eg} ... \sigma^{n_r}_{eg}\sigma^{n_1}_{ge} ...\sigma^{n_r}_{ge}$}.

\section{Power law decay for varying initial conditions}\label{sec:moreplaw}
\begin{figure}[h!]
\includegraphics[scale=0.5]{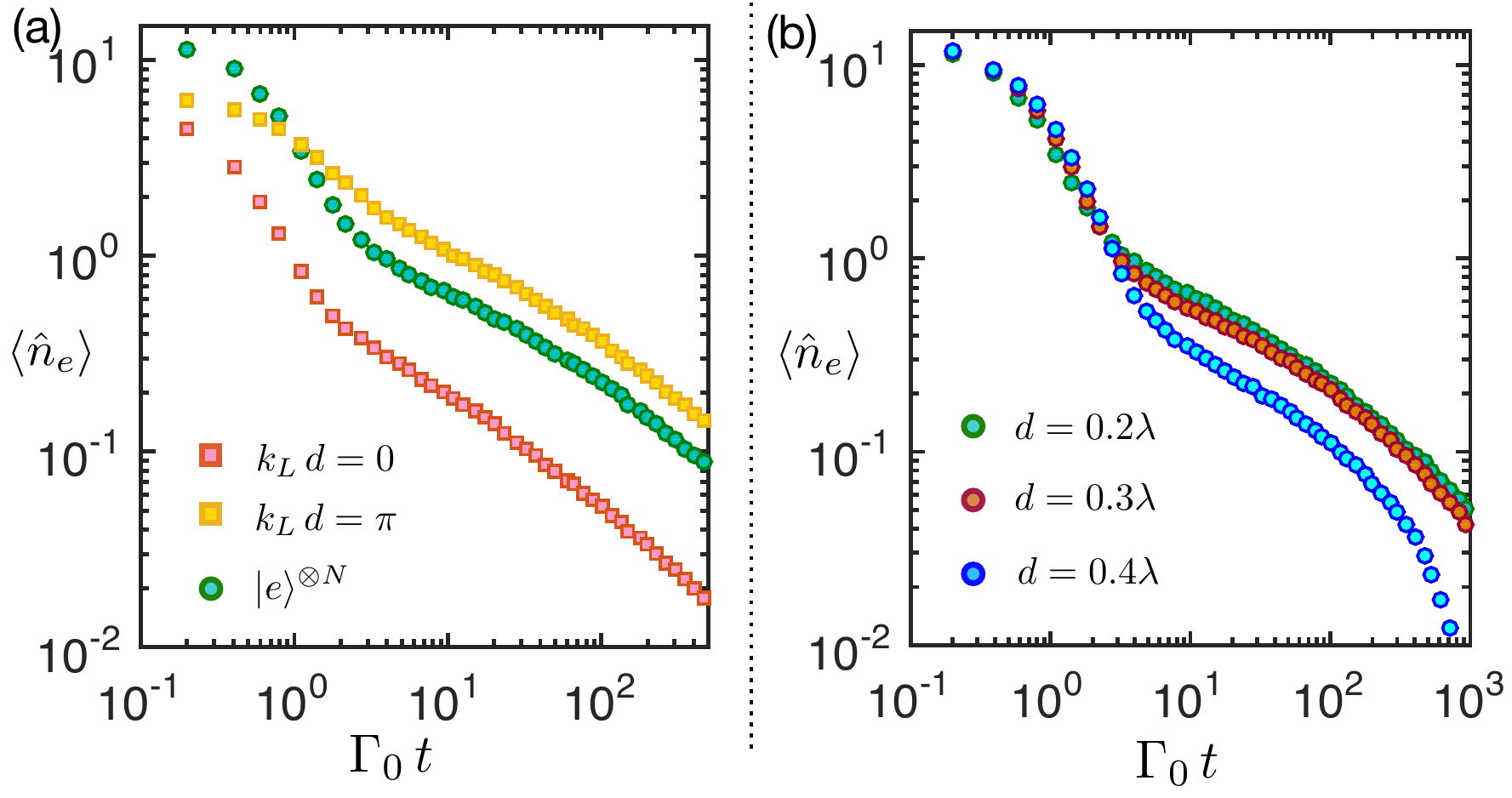}
\caption{\label{fig_plaw_init} Population decay for a free-space chain of $N=14$ atoms and (a) different initial states and a fixed inter-atomic distance $d=0.2\lambda_0$ and (b) the initial state $\ket{\psi(0)}=\ket{e}^{\otimes N}$ and various distances $d$. The initial states in (a) are either given by the totally excited state $\ket{e}^{\otimes N}$ (circle) or a clock state with wavevector $k_L$ as indicated in the figure (squares).}
\end{figure}

In Sect.\,\ref{sec:power_law_observation} a power-law in the population decay dynamics has been identified for an initially fully excited state $\ket{e}^{\otimes N}$ and a free space atomic chain of inter-atomic distance $d=0.2\lambda_0$. Here, we provide further evidence that the overall decay features are robust to both the initial decay and the specific chain parameters. Fig.\,\ref{fig_plaw_init}\,(a) depicts the excitation decay for clock states $\ket{\psi(0)}=\bigotimes_n (\ket{g_n}+e^{i k_L d\,n}\ket{e_n})/\sqrt{2}$, as introduced in Sect.\,\ref{sec:clock}, of wavevectors $k_L\,d=0$ and $k_L\,d=\pi$, respectively. For comparison the decay of a fully excited state equivalent to the one in Fig.\,\ref{Fig2} is shown. A qualitatively similar decay behavior is found for all initial states, consisting of a fast decay followed by a power-law region of similar scaling. The overall population in the long time limit depends on the portion of subradiant components. That is, states of more subradiant wavevectors ($k_L\,d=\pi$) retain higher populations than states dominantly involving radiant components ($k_L\,d=0$). The decay for an initially fully excited state and different inter-atomic distances $d$ is shown in Fig.\,\ref{fig_plaw_init}\,(b). Again the same decay characteristics {hold} true for all configurations, {with smaller lattice constants $d$ leading to larger long-time populations due to the increased presence of subradiant states.}

\section{Rate model for the eigenstate populations}\label{append_rmod}
In Sect.\,\ref{sec_jumppic} of the main text we introduced a semi-classical rate model for the decay dynamics, based on the insight that coherences play a minor role in the population dynamics. Here we compare the results obtained that way  to the ones obtained by solving the spin-model master equation. 

\begin{figure}[h!]
\includegraphics[scale=0.45]{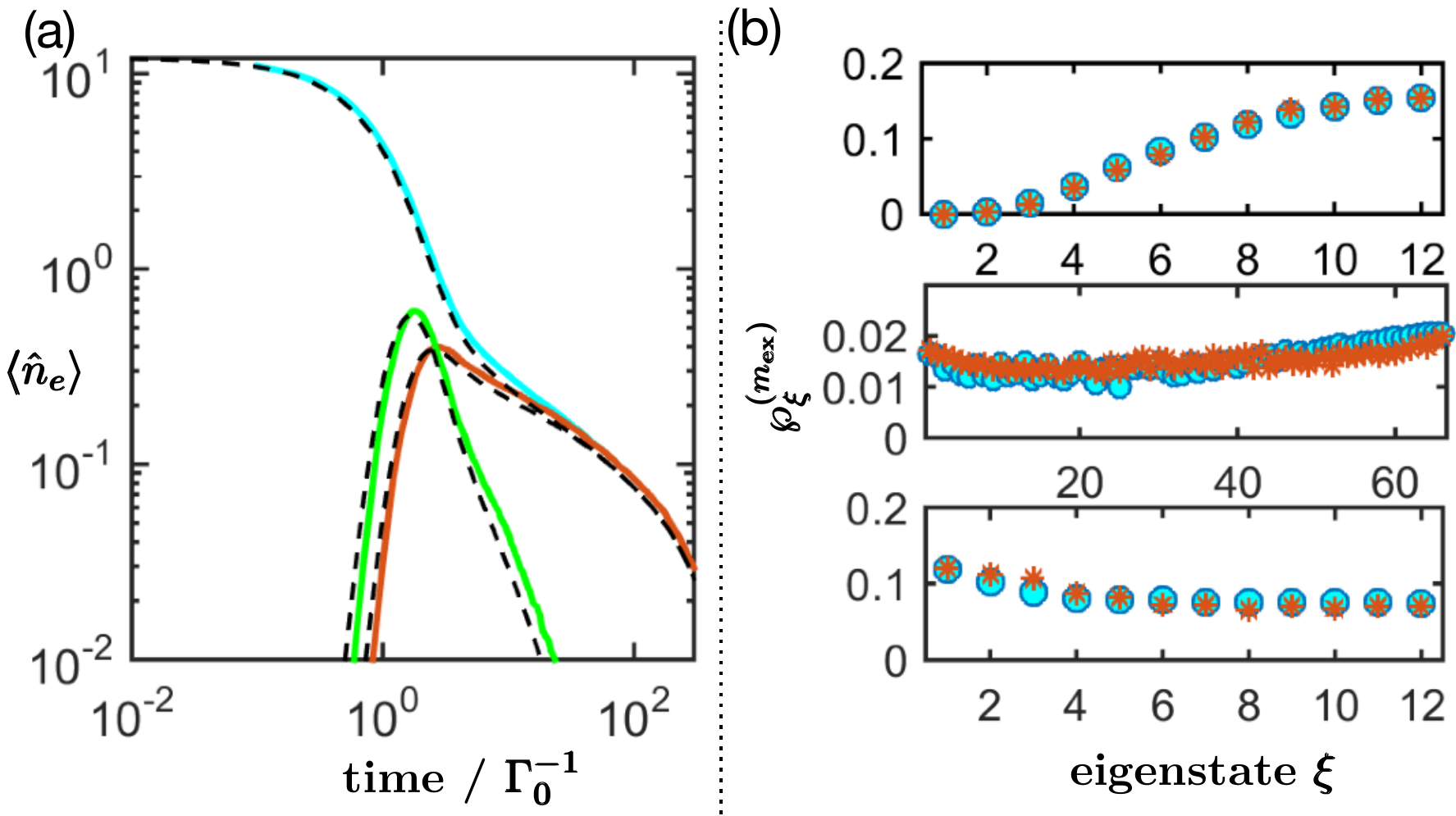}
\caption{\label{figSMrate} [free space atomic chain, $N=12$ atoms, $d=0.4\,\lambda_0$ and dipole orientation along the chain] (a) Excited state population $\ex{\hat{n}_e}$ decay of an initially fully excited state $\ket{e}^{\otimes N}$. Blue, red and green solid lines correspond to the total, single-excitation and two-excitation manifold population, respectively, and are obtained by solving the spin-model master equation. Black dashed lines have been calculated based on the rate equation. (b) Probability for passing through eigenstates for the excitation manifolds (from top to bottom) $m_{\rm ex}=11$, $m_{\rm ex}=2$ and $m_{\rm ex}=1$. Blue circles are based on the rate model (Eq.\,(\ref{eq_rmodel2}) in the main text), red stars are obtained from a quantum trajectory simulation of the master equation.  }
\end{figure}

Fig.\,\ref{figSMrate}\,(a) shows the excitation decay in time for an atom chain in free space, calculated by solving the master equation (solid lines) or by solving the coupled equations of the rate model ({Eq.}\,(\ref{eq_rmodel1}) in the main text, dashed lines). Good agreement is found between both approaches.

Fig.\,\ref{figSMrate}\,(b) illustrates the probability for passing through eigenstates for selected excitation manifolds. Blue circles are calculated based on the rate model and correspond to the ones in Fig.\,\ref{Fig3} in the main text. Red stars have been calculated by solving the spin-model master equation in the quantum trajectory method\,\cite{Daley14}: after each jump (decay) event the overlap probabilities of the resulting state with the eigenstates are recorded, from which a distribution is obtained by averaging over many (here: $10^3$) trajectories. Again, both approaches are in good agreement.

\section{\label{appendix:MPS} MPS simulation of the decay dynamics}
The exponentially growing Hilbert space with atom number $N$  renders  its full simulation for $N\gtrsim 20$ intractable in practice. Matrix Product States (MPS) provide an efficient way to approximate states in a much smaller subspace, provided that the correlations (entanglement)  between atoms remain limited. In reference\,\cite{manzoni17} it has been shown that the spin-model Hamiltonian, or more specifically the 1D waveguide Hamiltonian Eq.\,(\ref{eq:Effective_Hamiltonian_wg}), can be efficiently expressed and simulated in the MPS framework. We used that insight for the simulation of $N=30$ atoms in Fig.\,\ref{Fig2}. Here, we give details on the MPS simulation procedure. 

In general, a quantum state of $N$ atoms can be expressed in MPS form as
\begin{equation}\label{MPS1}  \ket{\psi} = \sum_{\sigma_1,\sigma_2,\dots\sigma_N}{A}^{\sigma_1}A^{\sigma_2}\dots A^{\sigma_N}\,\ket{\sigma_1,\sigma_2,\dots,\sigma_N}  \end{equation}
where $\sigma_k$ represents the local states of atom $k$ (e.g., the excited state $e$ or ground state $g$) and $A^{\sigma_k}$ are matrices that depend on the state $\sigma_{k}$. That is, the amplitude of the basis state $\ket{\sigma_{1},...,\sigma_{N}}$ is expressed as a ``matrix product''. The maximum dimension of matrices $A^{\sigma_k}$ - the maximum bond dimension - grows exponentially with the atom number, and the (efficient) approximate nature comes in by the restriction to a maximum bond dimension $D$. In the same way, operators are conveniently expressed in matrix product operator (MPO) form 
\begin{equation} \hat{O}=\prod_n \hat{W}^{[n]} =\prod_n \sum_{\sigma_n,\sigma_n'}W^{\sigma_n,\sigma_n'}\,\ket{\sigma_n}\bra{\sigma_n'}  \end{equation}
where again $W^{\sigma_n,\sigma_n'}$ for fixed $\sigma_n$ and $\sigma_{n'}$ form matrices. That way, the new MPS matrices $A'^{\sigma_k}$ of the state $\ket{\psi'}=\hat{O}\ket{\psi}$ can be straightforwardly obtained by contracting $W^{\sigma_k,\sigma_k'}$ with ${A}^{\sigma_k}$.  

One possible way to approximately solve the master equation (\ref{eq:Liouville_master_equation}) with the waveguide Hamiltonian (\ref{eq:Effective_Hamiltonian_wg}) is to ``vectorize" the density matrix $\rho$ and represent it in MPS form\,\cite{verstraete04}. In particular, we transform the density matrix to a wavevector representation $\hat{\rho} = \sum_{m,n}\rho_{m,n}\,\ket{m}\bra{n}$ $\rightarrow$ $\ket{\rho}=\sum_{m,n}\rho_{m,n}\ket{m}\otimes\ket{n}$. That way the density matrix can be written in the form (\ref{MPS1}) with a four-state local basis $\sigma_k\in\{\ket{e}\otimes{\ket{e}},\,\ket{g}\otimes{\ket{e}},\,\ket{e}\otimes{\ket{g}},\ket{g}\otimes{\ket{g}} \}$. Moreover, the Liouvillian $\mathrm{d}\ket{\rho}/\mathrm{d}t=\mathcal{L}\ket{\rho}$ takes on the form
\begin{widetext}
\begin{equation}\label{MPS3}\begin{split} \mathcal{L}&=\frac{\Gamma_{0}}{2}\sum_{m>n}  \lambda^{m-n}\,\Bigl\{ \left[(\mathds{1}\otimes\sigma_{ge}^n)-(\sigma_{eg}^n\otimes\mathds{1}) \right] (\sigma_{ge}^m\otimes\mathds{1})+(\sigma_{ge}^n\otimes\mathds{1})\,\left[(\mathds{1}\otimes\sigma_{ge}^m-\sigma_{eg}^m\otimes\mathds{1})\right]\Bigr\} \\
&+\frac{\Gamma_0}{2}\sum_{m>n}\lambda^{*(m-n)}\Bigl\{\left[ (\sigma_{ge}^n\otimes\mathds{1})-(\mathds{1}\otimes\sigma_{eg}^n)  \right]\,(\mathds{1}\otimes\sigma_{ge}^m)+(\mathds{1}\otimes\sigma_{ge}^n)\,\left[(\sigma_{ge}^m\otimes\mathds{1})-(\mathds{1}\otimes\sigma_{eg}^m) \right]  \Bigr\}\,+\sum_n\hat{V}_n
\end{split}\end{equation}
where we defined $\lambda=e^{i\,kd}$ and $\hat{V}_n=-(i\omega_0+\Gamma_0/2)(\sigma_{ee}^n\otimes\mathds{1})+(i\omega_0-\Gamma_0/2)(\mathds{1}\otimes\sigma_{ee}^n)+\Gamma_0\,(\sigma_{ge}^n\otimes\sigma_{ge}^n)$.
From (\ref{MPS3}) the MPO matrices of the Liouvillian can be constructed as\,\cite{schollwoeck11}

\begin{equation}\label{MPS4} \hat{W}^{[n]}=\begin{pmatrix}\mathds{1}_n\otimes\mathds{1}_n &\begin{smallmatrix}\frac{\Gamma_0}{2}\lambda  \bigl[(\mathds{1}_n\otimes\sigma_{ge}^n)\\\qquad-(\sigma_{eg}^n\otimes\mathds{1}_n) \bigr]\end{smallmatrix}  & \frac{\Gamma_0}{2}\lambda (\sigma_{ge}^n\otimes\mathds{1}_n)  & \frac{\Gamma_0}{2}\lambda^* (\mathds{1}_n\otimes\sigma_{ge}^n) & \begin{smallmatrix}\frac{\Gamma_0}{2}\lambda^*\bigl[ (\sigma_{ge}^n\otimes\mathds{1}_n)\\\qquad-(\mathds{1}_n\otimes\sigma_{eg}^n)  \bigr] \end{smallmatrix} & \hat{V}_n\\[3ex]
0 & \lambda (\mathds{1}_n\otimes\mathds{1}_n) & 0 & 0 & 0 & \sigma_{ge}^n\otimes \mathds{1}_n\\
0 & 0 &\lambda(\mathds{1}_n\otimes\mathds{1}_n) & 0 & 0 & (\mathds{1}_n\otimes\sigma_{ge}^n)-(\sigma_{eg}^n\otimes\mathds{1}_n)\\
0 & 0 & 0 & \lambda^*(\mathds{1}_n\otimes\mathds{1}_n) & 0 & (\sigma_{ge}^n\otimes \mathds{1}_n)-(\mathds{1}_n\otimes\sigma_{eg}^n)\\
0 & 0 & 0 & 0 & \lambda^*(\mathds{1}_n\otimes\mathds{1}_n) & \mathds{1}_n\otimes\sigma_{ge}^n\\
0 & 0 & 0 & 0 & 0 & \mathds{1}_n\otimes\mathds{1}_n
\end{pmatrix}\,\end{equation}
\end{widetext}
with  special forms for $\hat{W}^{[1]}$ and $\hat{W}^{[N]}$, which are of vector form and only consist of the first row and last column, respectively. Such a compact form of the MPO is not known for the free-space Hamiltonian Eq.\,(\ref{eq:Effective_Hamiltonian}), which prevents a straightforward application of the MPS formalism to that case.

The time evolution is performed by calculating time steps $\ket{\rho(t+\mathrm{d}t)}=(\mathds{1}+\mathcal{L}\,\mathrm{d}t)\,\ket{\rho(t)}$. The MPO of the evolution operator $(\mathds{1}+\mathcal{L}\,\mathrm{d}t)$ directly follows out of (\ref{MPS4}) by simply replacing $\hat{V}_1\to \hat{V}_1+(\mathds{1}_1\otimes\mathds{1}_1)$ in $\hat{W}^{[1]}$ and replacing $\Gamma_0\to \Gamma_0\,\mathrm{d}t$ and $\hat{V}_n\to\hat{V}_n\,\mathrm{d}t$ in all $\hat{W}^{[n]}$. Subsequent to the application of the MPO, which increases the bond dimension, the MPS is compressed by variational compression back to its original dimension\,\cite{schollwoeck11}. The expectation value of an operator $\hat{O}^\dagger$ follows as $\ex{\hat{O}^\dagger}=\text{tr}(\hat{O}^\dagger\rho)=\ex{O|\rho}$, where $\ket{O}$ is the operator $\hat{O}$ in vector representation analogue to $\ket{\rho}$. 

\tblue{
\section{\label{appendix:perpendicular} Clock signal decay for a chain of atoms with polarization perpendicular to the chain axis}}

\begin{figure}[h!]
\includegraphics[scale=0.4]{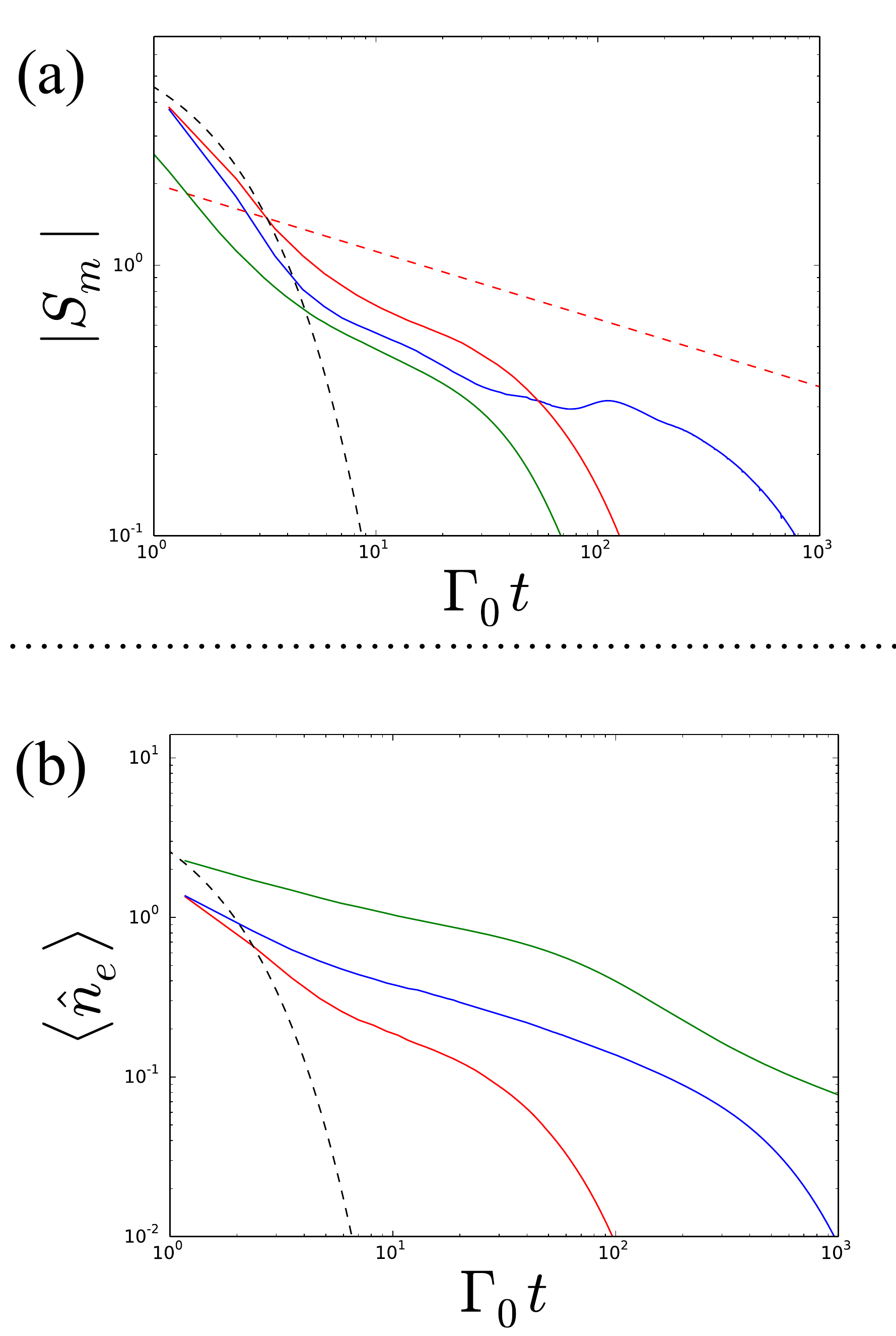}
\caption{\label{fig:signal_perp} \tblue{ (a) Evolution of the maximal clock signal $|S_m|$ in a free space configuration with the polarization of the atoms perpendicular to the chain axis, starting from the initial wavefunction  $\bigotimes_{n=1}^N \frac{|g_{n} \rangle+e^{in\pi} |e_{n}\rangle  }{\sqrt{2}}$,  $N=14$ atoms and inter-atomic distance $d/\lambda_0=0.4$ (in red), $d/\lambda_0=0.3$ (in blue), and $d/\lambda_0=0.2$ (in green). These results are calculated from an evolution of the wave function under stochastic quantum jumps. The {dashed black curve} shows the time-evolution of the signal amplitude for independent atoms, and the dashed red line shows a power-law guide to the eye with exponent $\nu=0.25$. (b) Evolution of the excitation number $\langle \hat{n}_e \rangle$ for the same protocol and parameters as in (a). The dashed black curve shows the dynamics of $\langle \hat{n}_e \rangle$ for independent atoms. }  }
\end{figure}

\begin{figure}[h!]
\includegraphics[scale=0.34]{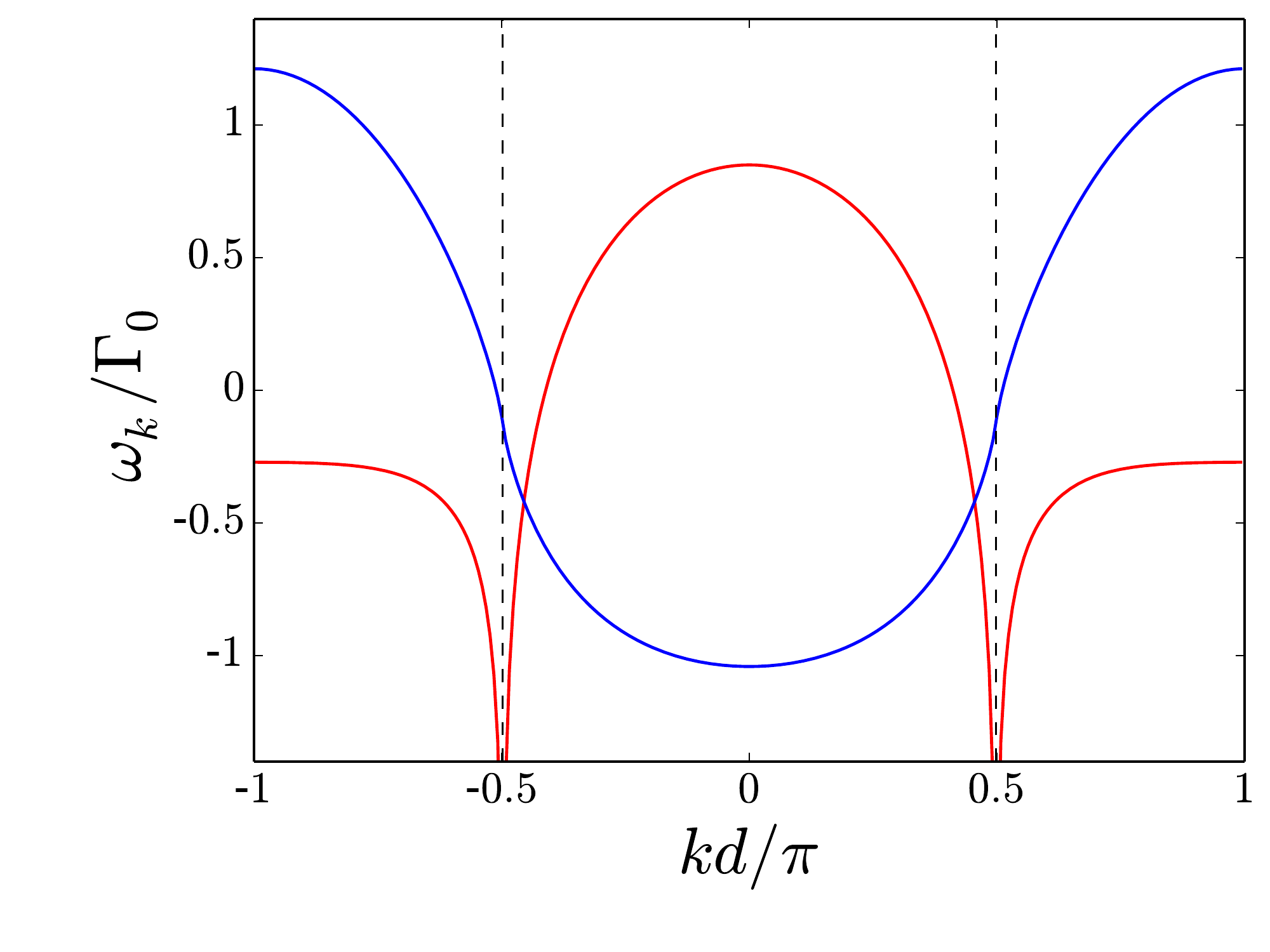}
\caption{\label{fig:dispersion_relation} \tblue{Single-excitation eigenstate energy shifts $\omega_k$ (normalized by $\Gamma_0$), as a function of wavevector $k$ for an infinite chain of atoms with atomic polarization parallel (in blue) and perpendicular (in red) to the chain axis\,\cite{asenjo17}, with $d/\lambda_0=0.25$. The dashed black lines correspond to $k=\pm k_0$.  }}
\end{figure}

\tblue{In Sec.\,\ref{sec:clock}, we have studied the clock dynamics for an atomic chain in free space with the atomic polarization parallel to the chain axis. Here, we focus on the case where the polarization of the atoms is perpendicular to the chain. In that case, the matrix elements in the effective Hamiltonian\,(\ref{eq:Effective_Hamiltonian}) read  ${\bf{p}}^\dagger \doublearrow{{\bf{G}}}({\bf{r}}_n,{\bf{r}}_m,\omega_0){\bf{p}}=(k_0^2 r^2+ik_0 r-1) |{\bf{p}}|^2 e^{i k_0 r}/(4\pi k_0^2 r^3)$. In Fig.\,\ref{fig:signal_perp}(a), we show the corresponding time-evolution of the maximal clock signal $|S_m|$, for an atomic chain of $N = 14$ atoms, and for selected values of $d/\lambda_0$. One observes regions of evolution where the decay appears close to a power law, with an exponent close to $\nu \simeq 0.25$. We interpret the overall longer persistence of the clock signal, as compared to the case of atomic parallel polarization, as arising from a smaller dephasing effect from coherent dipole-dipole interactions. This is illustrated in Fig.\,\ref{fig:dispersion_relation}, where we show the energy shifts $\omega_k$ of the single-excitation eigenstates of the effective Hamiltonian\,(\ref{eq:Effective_Hamiltonian}) as a function of their wavevector $k$ in the first Brillouin zone\,\cite{asenjo17}, both for an atomic chain with atomic polarization parallel and perpendicular to the chain axis. This quantity, and its dependence on $k$, quantifies the magnitude of coherent interactions involved in the eigenstates dynamics.} 

\tblue{ In Fig.\,\ref{fig:dispersion_relation}, it can be seen that the case of perpendicular atomic polarization yields an extremely flat band near the band edges $|k|d\approx \pi$, where subradiant states lie. This implies that the differential energy shifts between eigenstates are minimal, which would result in reduced many-body dephasing and is consistent with the observations of Fig.\,\ref{fig:signal_perp}(a).   }

\pagebreak
\bibliography{refs}

\end{document}